\documentclass[a4paper,11pt]{article}
\usepackage[no-natbib-sort]{jheppub} % for details on the use of the package, please
\usepackage{slashed} %Problem-Leo
\usepackage[T1]{fontenc} % if needed Problems-Leo
\usepackage{amsmath} 
\usepackage{mathtools}

\def\cN{\mathcal{N}}

\newcommand{\be}{\begin{equation}}
\newcommand{\ee}{\end{equation}}
\def\bea{\begin{eqnarray}}
\def\eea{\end{eqnarray}}

\newcommand{\matht}[1]{\ensuremath{\boldsymbol{#1}}}
\newcommand{\nn}{\nonumber}

\def\a{\alpha}

\def\CA{{\cal A}}

\def\CC{{\cal C}}

\def\CI{{\cal I}}

\def\CK{{\cal K}}

\def\CN{{\cal N}}
\def\CO{{\cal O}}

\def\CS{{\cal S}}
\def\CT{{\cal T}}

\def\CW{{\cal W}}

\def\BZ{{\mathbb Z}}

\def\beq#1\eeq{\begin{align}#1\end{align}}

\title{\boldmath Rotating Black Hole Entropy from M5-branes}

\author[a,b,c]{Francesco Benini,}
\author[d,e]{Dongmin Gang,}
\author[c,f]{and Leopoldo A. Pando Zayas}

\affiliation[a] {SISSA, Via Bonomea 265, 34136 Trieste, Italy}
\affiliation[b] {INFN, Sezione di Trieste, Via Valerio 2, 34127 Trieste, Italy}
\affiliation[c] {ICTP, Strada Costiera 11, 34151 Trieste, Italy}
\affiliation[d]{Quantum Universe Center, Korea Institute for Advanced Study, Seoul 02455, Korea}
\affiliation[e]{Asia Pacific Center for Theoretical Physics (APCTP), \\
Headquarters San 31, Hyoja-dong, Nam-gu, Pohang 790-784, Korea}
\affiliation[f]{Leinweber Center for Theoretical Physics, Randall Laboratory of Physics, \\
	The University of Michigan, Ann Arbor, MI 48109-1040, USA}

\abstract{We compute the superconformal index of 3d ${\cal N}=2$ superconformal field theories obtained from $N$ M5-branes wrapped on a hyperbolic 3-manifold. Exploiting the  3d-3d correspondence, we use perturbative invariants of $SL(N,\mathbb{C})$ Chern-Simons theory to determine the superconformal index in the large $N$ limit, including corrections logarithmic in $N$. The leading order partition function provides a microscopic foundation for the entropy function of the dual rotating asymptotically AdS$_4$ black holes. We also verify that the supergravity one-loop contribution to the $\log N$ term coincides with the field theoretic result. We propose a 3d-3d formulation for the refined topologically twisted index, and provide strong evidence in support of its vanishing --- which agrees with the fact that the expected dual rotating magnetically-charged black hole does not exist. This provides an interesting link between gravity and a  tantalizing mathematical  result.}

\begin{document} 
\begin{flushright}
{\tt\normalsize SISSA  27/2019/FISI \\ LCTP-19-24}\\
\end{flushright}
\setcounter{tocdepth}{2}
\maketitle

%%%%%%%%%%%%%%%%%%%%%%%%%%%%%%%%%%%%%%%%%%%%%%

\section{Introduction}

One of the most basic quantum properties of black holes is that they have an entropy \cite{Bekenstein:1972tm, Bekenstein:1973ur, Bekenstein:1974ax, Hawking:1974rv, Hawking:1974sw}. Bekenstein and Hawking understood that at leading order, the entropy in Einstein gravity is fixed in terms of the horizon area by the simple formula $S_\text{BH} = \mathrm{Area}/4G_\text{N}$ where $G_\text{N}$ is the Newton constant. This formula is universal: it applies in all dimensions, and for all types of smooth black holes. One of the great successes of string theory was to reproduce the Bekenstein-Hawking entropy of certain asymptotically-flat BPS black holes through a microscopic computation \cite{Strominger:1996sh}.

In the context of asymptotically AdS$_4$ black holes, string theory in its AdS/CFT guise has again provided a microscopic explanation for the Bekenstein-Hawking entropy \cite{Benini:2015eyy, Benini:2016rke}. Namely, it has been shown that the topologically twisted index \cite{Benini:2015noa} of a certain field theory precisely matches the entropy function (\textit{i.e.}, the grand canonical partition function) of the dual magnetically-charged  asymptotically AdS$_4$ black holes. Similar microscopic explanations have now been extended to a wide range of contexts, \textit{e.g.}, in four \cite{Hosseini:2016tor, Cabo-Bizet:2017jsl, Azzurli:2017kxo, Hosseini:2017fjo, Benini:2017oxt}, five \cite{Hosseini:2016cyf, Cabo-Bizet:2018ehj, Choi:2018hmj, Benini:2018ywd}, six \cite{Hosseini:2018uzp, Crichigno:2018adf, Suh:2018tul, Hosseini:2018usu, Suh:2018szn, Suh:2018qyv, Fluder:2019szh} and seven \cite{Kantor:2019lfo} dimensions (see the review \cite{Zaffaroni:2019dhb} for a more complete list of references).

For  rotating electrically-charged asymptotically AdS$_4$ black holes an impressive result was reported based on field theoretic computations of the superconformal  index \cite{Choi:2019zpz}; more recently an explanation based on localization was provided in \cite{Nian:2019pxj}. In this manuscript our goal is to approach a class of rotating, electrically-charged asymptotically AdS$_4$ black holes via their dual field theory.  We concentrate on configurations that are related via dual descriptions to a stack of  $N$ M5-branes.  An advantage of this approach is that it allows us to make use of the 3d-3d correspondence \cite{Dimofte:2011ju, Dimofte:2011py}. Recall that the 3d-3d correspondence states an equivalence between certain 3d ${\cal N}=2$ superconformal field theories $\CT_N[M_3]$ and Chern-Simons theory with gauge group $SL(N,\mathbb{C})$ on hyperbolic three-manifolds $M_3$.  This correspondence has the potential to provide an expression, exact in $N$, for various partition functions of the field theories through known Chern-Simons results. Most of the field theoretic approaches to partition functions or indices ultimately reduce the problem to a matrix model, whose exact solution has not been established yet. The exactness of the 3d-3d correspondence becomes therefore a crucial advantage, particularly in the analysis of subleading orders. 

More precisely, in this manuscript we present a computation of the superconformal index of the field theories, and show that it produces the entropy functions that generate the entropy of dual rotating asymptotically AdS$_4$ black holes through a Legendre transform.  We  also compute the one-loop effective action in 11d supergravity for M5-branes wrapped on the hyperbolic 3-space. We then match the gravity result with the field theory answer. In particular, we demonstrate that the agreement persists even when we consider hyperbolic 3-manifolds that are more general than those previously considered in the literature, including non-trivial first Betti number $b_1$. Our subleading computation has the potential to distinguish between various supersymmetric observables that have been shown to yield the same entropy function at leading order. 

Finally, we use the 3d-3d correspondence to evaluate the refined topologically twisted index, and find that it vanishes in the large $N$ limit. This matches with the fact that there are no BPS rotating magnetically-charged black holes with spherical horizon topology in 4d $\cN=2$ minimal gauged supergravity. In fact, we provide evidence of an even stronger result: that the refined topologically twisted index exactly vanishes at finite $N$. This result is a mathematical curiosity that has been previously entertained. It translates to a very non-trivial property of Chern-Simons perturbative invariants of hyperbolic three-manifolds, that would be interesting to mathematically explore.%
\footnote{Some work in this direction is in progress \cite{Gang:2019dbv}.}
We will offer a physical argument for such a vanishing.

The rest of the paper is organized as follows.
In section~\ref{Sec:Indices} we provide a 3d-3d perspective on various indices computed in the corresponding 3d ${\cal N}=2$ superconformal field theories.
In section~\ref{Sec:LargeN} we compute the indices in the large $N$ limit.  Section~\ref{Sec:Sugra} discusses properties of the supergravity description of wrapped M5-branes on hyperbolic 3-manifolds, including some of the relevant solutions.
Some details regarding the one-loop correction to terms logarithmic in $N$, from the supergravity point of view, are presented in section~\ref{Sec:OneLoop}.
We conclude in section~\ref{Sec:Conclusions} where we also highlight a number of interesting problems.
In appendix~\ref{App:Examples} we provide explicit examples of state-integral models for complex Chern-Simons theory, from which perturbative Chern-Simons invariants can be computed. 

\bigskip

{\bf Note added:} While we were preparing this manuscript for submission, we received \cite{Bobev:2019zmz} which has some overlap regarding the leading behavior of the superconformal index.

%%%%%%%%%%%%%%%%%%%%%%%%%%%%%%%%%%%%%%%%%%%%%%%%%%%%%%%%%%%%%%

\section{3d-3d relations for 3d indices}
\label{Sec:Indices}

We begin by studying the 3d-3d relations for three types of 3d supersymmetric indices: the superconformal index, the refined topologically twisted index on $S^2$, and the topologically twisted index on a general Riemann surface $\Sigma_g$ of genus $g$.  
Schematically, the 3d-3d relations take the following form:
\begin{multline}
\label{form of 3d-3d relations}
\textrm{A supersymmetric index of the 3d $\mathcal{N}=2$ $\mathcal{T}_N[M_3]$ theory} {} \\
{} = \textrm{An invariant of $SL(N,\mathbb{C})$ Chern-Simons theory on $M_3$.}
\end{multline}
In subsections \ref{sec : TNM theory} and \ref{sec : SL(N) CS theory} we will explain the basic definitions and properties of the two quantum field theories --- a supersymmetric theory and a topological theory, respectively --- appearing in the relations. Then, in subsection~\ref{sec : 3d-3d relations} we will give precise formulations of the 3d-3d relations for the three types of indices.
The part on the superconformal index will basically be a review of previous work \cite{Dimofte:2011py, Lee:2013ida, Yagi:2013fda}, with some additional clarifications. On the other hand, we will propose a 3d-3d relation for the refined version of the twisted index, with non-trivial supporting evidence. Motivated by that, we will also propose a \mbox{3d-3d} relation for the twisted index on general Riemann surfaces, thus generalizing previous work \cite{Gang:2019uay} and covering general hyperbolic 3-manifolds $M_3$. Finally, in subsection~\ref{sec : Integrality check} we will provide non-trivial consistency checks for the proposed 3d-3d relations, confirming the expected integral properties of the various indices.

\subsection[{3d $\CT_N[M_3]$ theory}]{3d \matht{\CT_N[M_3]} theory}
\label{sec : TNM theory}

Let us first recall the definition and basic properties of the 3d $\mathcal{T}_N[M_3]$ theory appearing on the left-hand-side (LHS) of the 3d-3d relation \eqref{form of 3d-3d relations}. We define 
\begin{align}
\begin{split}
\mathcal{T}_N [M_3] & := \Big( \textrm{the 3d $\CN{=}2$ superconformal field theory obtained from a twisted}
\\[-.5em]
& \qquad\quad \textrm{compactification of the 6d $\mathcal{N}{=}(2,0)$ theory of type $A_{N-1}$ on $M_3$} \Big) \;,
\\[.3em]
M_3 &= \big( \textrm{a {\it closed hyperbolic} 3-manifold} \big)  = \mathbb{H}^3 /\Gamma \;. \label{3d-3d set-up}
\end{split}
\end{align}
We use an $SO(3)$ subgroup of the $SO(5)$ R-symmetry of the 6d theory to perform partial topological twist along $M_3$. The resulting 3d theory $\CT_N[M_3]$ has 3d $\CN=2$ supersymmetry, with no (non-R) flavor symmetries at sufficiently large $N$.  The $U(1) \cong SO(2)$ R-symmetry of the 3d theory comes from the $SO(2) \subset SO(2)\times SO(3) \subset SO(5)$ subgroup of the 6d R-symmetry, and thus its charges $R$ are quantized:
\begin{align}
R \in \mathbb{Z} \quad \textrm{  in $\CT_N[M_3]$ theory} \;. \label{quantization of R-charge}
\end{align}
At small $N$ there could be an accidental flavor symmetry in $\mathcal{T}_N[M_3]$ as studied in \cite{Gang:2017lsr, Gang:2018wek}. In that case, the infrared (IR) superconformal R-symmetry could be different from the R-symmetry originated from $SO(2) \subset SO(5)$. The field theoretic construction of $\mathcal{T}_N[M_3]$ was proposed in \cite{Dimofte:2011ju, Dimofte:2013iv, Gang:2018wek} based on a Dehn-surgery representation of $M_3$ using a link $L \subset S^3$ and an ideal triangulation of the  link complement $S^3\backslash L$. The theory $\mathcal{T}_N[M_3]$  is expected to be independent of the choice of Dehn-surgery representation and ideal triangulation.%
\footnote{This approach requires that the link complement be hyperbolic.}
Such an invariance gives a geometric understanding of certain dualities among 3d $\mathcal{N}=2$ gauge theories.

\subsection[$SL(N,\mathbb{C})$ Chern-Simons theory on $M_3$]{\matht{SL(N,\mathbb{C})} Chern-Simons theory on \matht{M_3}}
\label{sec : SL(N) CS theory}

The most general action for $SL(N,\mathbb{C})$ Chern-Simons theory on $M_3$ is  given by 
\begin{align}
S_{k, \sigma} [\CA, \widetilde{\CA};M_3] &= \frac{(k+i \sigma)}{8\pi} \int_{M_3} \textrm{Tr} \left( \CA \wedge d \CA + \frac{2}3 \CA^3 \right) + \frac{(k-i \sigma)}{8\pi} \int_{M_3} \textrm{Tr} \left( \widetilde{\CA} \wedge d \widetilde{\CA} + \frac{2}3 \widetilde{\CA}^3 \right)
\nn \\
& = \frac{2\pi i}{\hbar} \, \mathrm{CS}[\CA; M_3] + \frac{2\pi i}{ \tilde{\hbar}} \, \mathrm{CS}[\widetilde{\CA};M_3] \;.
\end{align}
We have introduced (anti-)holomorphic couplings and the Chern-Simons functional:
\begin{align}
\begin{split}
&\hbar  := \frac{4\pi i}{k+ i \sigma} \;, \qquad \tilde{\hbar}:=\frac{4\pi i}{k - i \sigma}\;,
\\
& \mathrm{CS}[\CA; M_3] := \frac{1}{4\pi}\int_{M_3} \textrm{Tr} \left(\CA \wedge d \CA + \frac{2}3 \CA^3 \right) \;.
\end{split}
\end{align}
For gauge invariance and  unitarity of the theory, it is required that 
\begin{align}
k \in \mathbb{Z} \;, \qquad\qquad \sigma \in \mathbb{R} \textrm{ or } i \mathbb{R}\;. 
\end{align}
The $SL(N,\mathbb{C})$ Chern-Simons partition function is defined by the following path-integral:
\begin{align}
\begin{split}
&\mathcal{Z}^{SL(N,\mathbb{C})}_{k, \sigma} \big[ M_3;\{n_{\alpha \beta}\} \big] = \int_{\Gamma( \{n_{\alpha \beta}\} )} \frac{[D \CA][D \widetilde{\CA}]}{(\textrm{gauge})}  \; e^{i S_{k,\sigma} [\CA, \widetilde{A};M_3]} \;,
\\
&\Gamma \big( \{n_{\alpha \beta}\} \big) := \sum_{\mathcal{A}^\alpha, \widetilde{\CA}^\beta \in \chi (M_3;N)} n_{\alpha,\beta} \; \CC^{\alpha}(\CA)\times \CC^{\beta} (\widetilde{\CA}) \;,
\end{split}
\end{align}
where $\chi (M_3;N)$ represents
\begin{align}
\chi(M_3;N) := \big( \textrm{set of gauge-inequivalent $SL(N,\mathbb{C})$ flat connections on $M_3$} \big) \;.
\end{align}
Moreover, $\CC^\alpha (\CA)$ denotes the absolutely convergent integration cycle (Lefschetz thimble) in the configuration space of gauge fields $\CA$, associated to a flat connection  $\CA^\alpha$. (See, for example, \cite{Witten:2010cx} for details on the integration cycle $\CC^{\alpha}(\CA)$.) Thus, to specify a quantum $SL(N,\mathbb{C})$ Chern-Simons theory, we need to specify the choice of a consistent integration cycle $\{n_{\alpha \beta} \}$ as well as the CS levels $(k,\sigma)$.

The partition function can be factorized as
\begin{align}
\mathcal{Z}^{SL(N,\mathbb{C})}_{k, \sigma} \big[ M_3;\{n_{\alpha \beta}\} \big]  = \sum_{\mathcal{A}^\alpha, \widetilde{\mathcal{A}}^\beta \in \chi(M_3;N) } \frac{1}{\bigl| \mathrm{Stab}(\alpha, \beta) \bigr|} \; n_{\alpha, \beta} \; B_{M_3}^{\alpha}(q;N) \; B_{M_3}^{\beta}(\tilde{q};N)\;.
\end{align}
Here the ``holomorphic blocks'' $B^{\alpha}(q)$ are defined as
\begin{align}
\label{CS holomprhic block}
B_{M_3}^{\alpha}(q:=e^\hbar;N) = \int_{\CC^{\alpha}(\CA)} \frac{[D\CA]}{(\rm gauge)} \; e^{-\tfrac{2\pi}{\hbar} \mathrm{CS}[\CA;M_3]} \;. \end{align}
We defined the exponentiated holomorphic and anti-holomorphic couplings as
\begin{align}
q:=e^{\hbar} = e^{\frac{4\pi i }{k +i \sigma}}\;, \qquad\quad \tilde{q} := e^{\tilde{\hbar}} = e^{\frac{4\pi i }{k - i \sigma}}\;.
\end{align}
The prefactor $\bigl| \mathrm{Stab}(\alpha, \beta) \bigr|$ is the volume of the stabilizer of $(\CA^\alpha, \widetilde\CA^\beta)$, \textit{i.e.}, the volume of the subgroup of the gauge group that preserves the flat connections.
In an asymptotic  $\hbar \rightarrow 0$ limit, the holomorphic blocks $B^\alpha$ can be perturbatively expanded:
\begin{align}
B^\alpha_{M_3} (q) \;\; \xrightarrow{\; \hbar \rightarrow 0 \;} \;\; \exp \left(\frac{1}\hbar S^{\alpha}_0(M_3;N) + S^{\alpha}_1 (M_3;N) + \ldots + \hbar^{n-1} S_n^{\alpha}(M_3;N) +\ldots \right) \;.
\end{align}
For example
\begin{align}
\label{Classical/1-loop}
S_0^\alpha = -2\pi \, \mathrm{CS}[\CA^\alpha;M_3]\;, \qquad\qquad S_1^\alpha = - \frac{1}2 \, \mathrm{Tor}[\CA^\alpha;M_3]\;,
\end{align}
where $\mathrm{Tor}[\CA^\alpha;M_3]$ is the analytic Ray-Singer torsion twisted by the flat connection $\CA^\alpha$ defined as follows \cite{ray1971r, gukov2008sl}:
\begin{align}
\mathrm{Tor}[\mathcal{A}^\alpha ;M_3] := \frac{\bigl[ \det ' \Delta_1 (\mathcal{A}^\alpha) \bigr]^{1/4}}{\bigl[ \det ' \Delta_0 (\mathcal{A}^\alpha) \bigr]^{3/4}} \;. \label{Def of Ray-Singer torsion}
\end{align}
Above, $\Delta_n (\mathcal{A}^\alpha)$ is the Laplacian acting on $\mathfrak{sl}(N,\mathbb{C})$-valued $n$-forms twisted by a flat connection  $\mathcal{A}^\alpha$:
\begin{align}
\Delta_n (\mathcal{A}) = d_\CA * d_\CA * \, + \, *d_\CA *d_\CA\;, \qquad\quad d_\CA  = d+ \mathcal{A}\wedge\;.
\end{align}
Then, $\det'\Delta$ denotes the zeta-function regularized product of non-zero eigenvalues of  $\Delta$. To define the Laplacian, one needs to introduce a  smooth metric  on $M_3$. The  torsion is however independent of the metric choice and is a topological invariant.

\subsection{3d-3d relations}
\label{sec : 3d-3d relations}

We  consider three types of  supersymmetric indices of 3d $\mathcal{N}=2$ gauge theories: the superconformal index, the refined topologically twisted index on $S^2$, and twisted index on general closed Riemann surfaces $\Sigma_g$ of genus $g$. Let us define
\begin{align}
\begin{split}
\CI_{\rm sci}\big( q;\CT_{N}[M_3] \big) &= \big( \textrm{Superconformal index of $\CT_N[M_3]$ theory on $S^2$} \big) \;
\\
&:=\textrm{Tr}_{\mathcal{H}_{\rm sci}(S^2)} \, (-1)^R \, q^{\frac{R}2 +j_3} \;,
\\[.8em]
\CI_{\rm top}\big( q;\CT_{N}[M_3] \big) &= \big( \textrm{Refined topologically twisted index of $\CT_N[M_3]$ theory on $S^2$} \big) \;
\\
&:=\textrm{Tr}_{\mathcal{H}_{\rm top}(S^2)} \, (-1)^R \, q^{j_3} \;,
\\[.8em]
\CI_{\Sigma_g}\big( \CT_{N}[M_3] \big) &= \big( \textrm{Topologically twisted index of $\CT_N[M_3]$ theory on $\Sigma_g$} \big) \;
\\
&:=\textrm{Tr}_{\mathcal{H}_{\rm top}(\Sigma_g)} \, (-1)^R  \;.
\end{split}
\end{align}
Here $R$ is the R-charge corresponding to the compact $SO(2)$ R-symmetry subgroup of $SO(5)$, while $j_3$ is the Cartan generator of the $\mathfrak{su}(2) \cong \mathfrak{so}(3)$ isometry of $S^2$. To produce protected indices, instead of $(-1)^F$ or $(-1)^{2j_3}$ we use $(-1)^R$  which typically appears in the context of 3d-3d correspondences.
Then $\mathcal{H}_{\rm sci}(S^2)$ is the radially-quantized Hilbert space of $\mathcal{T}_{N}[M_3]$, whose elements are linear combinations of local operators. Only $1/4$ BPS operators satisfying $\Delta = R +j_3$, where $\Delta$ is the conformal dimension, contribute to the index. 
On the other hand, $\mathcal{H}_{\rm top}(\Sigma_g)$ is the Hilbert space of  the topologically twisted  $\mathcal{T}_{N}[M_3]$ theory on a closed Riemann surface $\Sigma_g$. In the topological twisting, we turn on a background  gauge field $A_R^{\rm background} $ coupled to the $SO(2)$ R-symmetry along $\Sigma_g$ in such a way that
\begin{align}
A_R^{\rm background} = \frac12 \, \omega (\Sigma_g) \quad \Rightarrow  \quad \frac{1}{2\pi}\int_{\Sigma_g} d A_R^{\rm backgrond} =(1-g)\;.
\end{align}
Here $\omega (\Sigma_g)$ is the spin-connection on $\Sigma_g$. Due to the background magnetic flux, the following Dirac quantization conditions are required:
\begin{align}
(g-1) \, R (\mathcal{O}) \in \mathbb{Z}
\end{align}
for all scalar gauge-invariant operators $\CO$ in the theory. Thanks to \eqref{quantization of R-charge}, the quantization condition is automatically satisfied for $\mathcal{T}_N[M_3]$ theories and we can consider the twisted index for arbitrary $g$. When $g=0$, \textit{i.e.}, for $\Sigma_g = S^2$, there is an additional $\mathfrak{su}(2)$ rotational symmetry in the topologically twisted background, and we can refine the twisted index by turning on a fugacity $q$ associated to the Cartan generator $j_3$ of $\mathfrak{su}(2)$. As typical for a Witten index, the twisted index only gets contributions from ground states in $\mathcal{H}_{\rm top}(\Sigma_g)$, whose number is finite.

\subsubsection{Superconformal index}

The 3d-3d relation for the superconformal index was studied in \cite{Dimofte:2011py, Beem:2012mb, Yagi:2013fda, Gang:2013sqa, Lee:2013ida}, and the following proposal was made:
\begin{align}
\label{3d-3d for SCI-1}
\begin{split}
\CI_{\rm sci}\big( q;\CT_{N}[M_3] \big)  &=   \mathcal{Z}^{SL(N,\mathbb{C})}_{k=0 ,\, \sigma =\frac{2\pi}{\log q}} \left[M_3, \{n_{\alpha \beta} = \delta_{\alpha \overline{\beta}}\} \right]
\\
&= \sum_{\mathcal{A}^\alpha \in \chi(M_3;N) }  \frac{1}{|\mathrm{Stab}(\a)|} \, B_{M_3}^\alpha (q;N) \, B_{M_3}^{\overline{\alpha}} (q^{-1};N)
\\
&= \sum_{\mathcal{A}^\alpha   \in \chi_\mathrm{irred}(M_3;N)}  \frac{1}{|\mathrm{Stab}(\a)|} \, B_{M_3}^\alpha (q;N) \, B_{M_3}^{\overline{\alpha}} (q^{-1};N)\;.
\end{split}
\end{align}
For a reducible flat connection $\mathcal{A}^\alpha$, its stablizer subgroup is non-compact, \mbox{$\bigl| \mathrm{Stab}(\mathcal{A}^\alpha) \bigr| = \infty$} and thus it does not contribute to the path-integral --- see \cite{Chung:2014qpa, Dimofte:2016pua} for more detailed discussions on this issue. This is one crucial difference between Chern-Simons theory with compact and non-compact gauge group. Therefore, we only need to sum over a subset  $\chi_\mathrm{irred} (M_3;N)$ of flat connections defined as
\begin{align}
\chi_{\rm irred} (M_3;N) := \big( \textrm{set of {\it irreducible} gauge-inequivalent $SL(N,\mathbb{C})$ flat connections on $M_3$} \big) \;.
\end{align}
Alternatively, we can take the point of view advocated, \textit{e.g.}, in \cite{Chung:2014qpa, Gang:2018wek} that the theory $\CT_{N}[M_3]$ describes only part of the moduli space of $N$ M5-branes on $M_3$. This moduli sub-space is essentially a Higgs branch (as the theories $\CT_{N}[M_3]$ have Higgs but no Coulomb branch) and its points correspond the irreducible $SL(N,\mathbb{C})$ flat connections on $M_3$, leading to $\chi_\mathrm{irred}(M_3;N)$.

In (\ref{3d-3d for SCI-1}),  $\CA^{\overline{\a}}$ denotes  the complex conjugation of   $\CA^\alpha$:
\begin{align}
\CA^{\overline{\a}}  = (\CA^{\a})^*\;.
\end{align}
The 3d-3d relation \eqref{3d-3d for SCI-1} can be further simplified using the following fact:
\begin{align}
\label{Tensoring using center flat connection}
B^{\alpha_1}_{M_3} (q;N) = B^{\alpha_2}_{M_3} (q;N) \qquad
\begin{split} &\textrm{if $\mathcal{A}^{\alpha_1}$ and $\mathcal{A}^{\alpha_2}$ are related to each other by}
\\
&\textrm{tensoring with a $\mathbb{Z}_N $ flat-connection} \;.
\end{split}
\end{align}
Here $\mathbb{Z}_N$ is the center of $SL(N,\mathbb{C})$. Let us explain this tensoring more explicitly. Each $SL(N,\mathbb{C})$ flat-connection $\mathcal{A}^\alpha$ determines a group homomorphism $\rho^\alpha$
\begin{align}
\rho^\alpha \in \textrm{Hom}\big( \pi_1 (M_3) \rightarrow SL(N,\mathbb{C}) \big)/SL(N,\mathbb{C})\;
\end{align}
up to conjugation. Similarly, a $\mathbb{Z}_N$ flat-connection determines a homomorphism $\eta$
\begin{align}
\eta \in \textrm{Hom}\big( \pi_1 (M_3) \rightarrow \mathbb{Z}_N \big) = \textrm{Hom} \big( H_1(M_3,\mathbb{Z})\rightarrow \mathbb{Z}_N \big) \;.
\end{align}
By tensoring with $\eta$, we can obtain from a flat connection $\mathcal{A}^{\alpha}$ another flat connection $\mathcal{A}^{\eta \otimes \alpha}$  whose $SL(N,\mathbb{C})$ holonomies are given by 
\begin{align}
\label{1-form symmetry action}
\rho^{\eta \otimes  \alpha} (c) =\eta(c)  \cdot   \rho^\alpha (c)   \;, \qquad \forall c \in \pi_1 (M_3) \;. 
\end{align}
Here $\cdot$ is the group multiplication in $SL(N,\mathbb{C})$.
Since there are only adjoint fields in $SL(N,\mathbb{C})$  Chern-Simons theory, the equality in \eqref{Tensoring using center flat connection} holds. More precisely, (\ref{1-form symmetry action}) is the action of the $\mathbb{Z}_N$ 1-form center symmetry of $SL(N,\mathbb{C})$ Chern-Simons theory and it leaves the path-integrand invariant.
Using \eqref{Tensoring using center flat connection} and the fact that
\begin{align}
\begin{split}
&\Bigl{|} \mathrm{Stab}\bigl( \mathcal{A}^\alpha \in \chi_{\rm irred}(M_3;N) \bigr) \Bigr{|} = \Bigl{|}\textrm{Center of $SL(N,\mathbb{C})$}\Bigr{|} = N \;,
\end{split}
\end{align} 
we obtain the following final version of the 3d-3d relation for the superconformal index:
\begin{align}
\label{3d-3d for SCI}
\begin{split}
\CI_{\rm sci}\big( q;\CT_{N}[M_3] \big)  &=   \mathcal{Z}^{SL(N,\mathbb{C})}_{k=0 ,\, \sigma =\frac{2\pi}{\log q}} \left[M_3, \{n_{\alpha \beta} = \delta_{\alpha \overline{\beta}}\} \right]
\\[.5em]
&=   \frac{\bigl| \mathrm{Hom}[\pi_1 (M_3) \rightarrow \mathbb{Z}_N] \bigr|}{N}   \sum_{[\mathcal{A}^\alpha] \in \frac{\chi_{\rm irred} (M_3;N)}{\mathrm{Hom}[\pi_1 (M_3) \rightarrow \mathbb{Z}_N]} } B^\alpha_{M_3} (q;N) \; B^{\overline\alpha}_{M_3}(q^{-1};N) \;.
\end{split}
\end{align}
Here $[\mathcal{A}^\alpha]$ is an equivalence class in $\chi_{\rm irred}(M_3;N)$ under the equivalence relation defined by tensoring with $\mathbb{Z}_N$ flat connections:
\begin{align}
[\mathcal{A}^\alpha]= [\mathcal{A}^{\eta \otimes \alpha }] \;, \qquad \forall \eta \in \textrm{Hom}\bigl[ \pi_1 (M_3) \rightarrow \mathbb{Z}_N \bigr] \;.
\end{align}

\subsubsection{Refined twisted index}

From the factorization property of the refined twisted index \cite{Benini:2015noa, Nieri:2015yia},%
\footnote{The refined index of the tetrahedron ($\Delta$) theory $\mathcal{T}_{N=2}[M_3 = \Delta ]$ \cite{Dimofte:2011ju} can be written as $B_{\Delta} (q,z) \, B_{\Delta }(\tilde{q},\tilde{z})$ with $\tilde q = q^{-1}$ and $\tilde z = z$. In the 3d-3d relation, $q=e^{\frac{4\pi i }{k+ i \sigma}}$ and $\tilde{q}= e^{\frac{4\pi i }{k-i \sigma}}$.  So the relation $\tilde q = q^{-1}$ implies $k=0$. Then $z$ and $\tilde z$ parametrize boundary $SL(2,\mathbb{C})$ holonomies of the gauge fields $\mathcal{A}$ and $\widetilde{\mathcal{A}}$, respectively. Thus the relation $z = \tilde{z}$ implies $n_{\alpha \beta } = \delta_{\alpha \beta}$. From the point of view of the supersymmetric field theory, factorization of the partition function can be understood in terms of Higgs branch localization to vortex and anti-vortex partition functions \cite{Benini:2012ui, Benini:2013yva}.}
we naturally propose the following 3d-3d relation:
\begin{align}
\label{3d-3d for refined index}
\CI_{\rm top} \big( q;\CT_{N}[M_3] \big)  &=   \mathcal{Z}^{SL(N,\mathbb{C})}_{k=0 ,\, \sigma =\frac{2\pi}{\log q}} \bigl[ M_3, \{n_{\alpha \beta} = \delta_{\alpha\beta}\} \bigr]
\\[.5em]
&=   \frac{ \bigl| \mathrm{Hom}[\pi_1 (M_3) \rightarrow \mathbb{Z}_N] \bigr|}{N}   \sum_{[\mathcal{A}^\alpha] \in \frac{\chi_{\rm irred} (M_3;N)}{\mathrm{Hom}[\pi_1 (M_3) \rightarrow \mathbb{Z}_N]} }  B_{M_3}^\alpha (q;N) \; B^{\a}_{M_3}(q^{-1};N) \;.
\nn
\end{align}
This looks similar to (\ref{3d-3d for SCI}) but, as opposed to the former, this formula has no complex conjugation of the holomorphic blocks.
A study of the refined twisted index in the context of the 3d-3d correspondence already appeared in \cite{Gukov:2017kmk}. However, the relation we propose is different and complementary to theirs. In \cite{Gukov:2017kmk}, the twisted (half-)index was related to \emph{homological} blocks labelled by Abelian flat connections. On the other hand, we propose a relation with \emph{holomorphic} blocks labelled by irreducible flat connections.

As an initial consistency check of our proposal, we can obtain the 3d-3d relation for the (unrefined) twisted index at $g=0$ put forward in \cite{Gang:2019uay}, by taking the $q\to1$ limit of the refined index:
\begin{align}
\begin{split}
&\CI_\textrm{top}\big( q;\CT_{N}[M_3]\big) \Big{|}_{q\rightarrow 1}   
\\
&=   \frac{\bigl| \mathrm{Hom}[\pi_1 (M_3) \rightarrow \mathbb{Z}_N] \bigr|}{N}   \sum_{[\mathcal{A}^\alpha] \in \frac{\chi_{\rm irred} (M_3;N)}{\mathrm{Hom}[\pi_1 (M_3) \rightarrow \mathbb{Z}_N]} } B_{M_3}^\alpha (q;N)  \; B_{M_3}^{\a}(q^{-1};N) \Big{|}_{q\rightarrow 1} 
\\
& =   \frac{\bigl| \mathrm{Hom}[\pi_1 (M_3) \rightarrow \mathbb{Z}_N] \bigr|}{N}   \sum_{[\mathcal{A}^\alpha]}
\exp \left(\sum_{n=0}^\infty \hbar^{n-1} S_n^{\alpha} \right)  \exp \left( \sum_{n=0}^\infty (-\hbar)^{n-1} S_n^{\alpha} \right) \bigg{|}_{\hbar \rightarrow 0}
\\
&=   \frac{\bigl| \mathrm{Hom}[\pi_1 (M_3) \rightarrow \mathbb{Z}_N] \bigr| }{N}   \sum_{[\mathcal{A}^\alpha]}
\exp \bigl( 2 S_1^{\alpha}(M_3;N) \bigr) \;. 
\label{refined index at q=1}
\end{split}
\end{align}
Notice that only the 1-loop part contributes in the unrefined limit. The 3d-3d relation for twisted indices proposed in \cite{Gang:2019uay} is
\begin{align}
\label{3d-3d for twisted index}
\begin{split}
&\mathcal{I}_{\Sigma_g} \big( \mathcal{T}_N[M_3] \big) = \sum_{\alpha \in\chi_\mathrm{irred} (M_3;N) } \bigl(N \times  \exp (-2S_1^\alpha) \bigr)^{g-1} \;,
\\
& \textrm{for $M_3$ with trivial $H_1 (M_3, \mathbb{Z}_N)$} \;.
\end{split}
\end{align} 
Such a relation was indirectly derived by combining several recent technical developments: i) field theoretic constructions of $\mathcal{T}_N[M_3]$ \cite{Dimofte:2011ju, Dimofte:2013iv, Gang:2018wek}; ii) field theory localization formulas of twisted indices \cite{Benini:2015noa, Benini:2016hjo, Closset:2016arn}; iii) mathematical tools \cite{Dimofte:2009yn, Dimofte:2012qj, Gang:2017cwq} for computing the Ray-Singer torsion $-2S_1^\alpha$ from state-integral models for $SL(N,\mathbb{C})$ Chern-Simons theory. 
The 3d-3d relation (\ref{3d-3d for twisted index}) with $g=0$ precisely matches the  $q\rightarrow 1$ limit \eqref{refined index at q=1} of the proposed 3d-3d relation \eqref{3d-3d for refined index}.

\subsubsection{Twisted indices}

Let us compare the 3d-3d relation \eqref{3d-3d for refined index} with the following universal expression for the twisted index $\mathcal{I}_{\Sigma_{g=0}}$ of general 3d $\mathcal{N}=2$ theories $\mathcal{T}$:
\begin{align}
\mathcal{I}_{\Sigma_{g=0}} (\mathcal{T}) =  \sum_{\alpha \in \mathcal{S}_\mathrm{BE} (\mathcal{T})}  (\mathcal{H}_\alpha)^{-1} \;.
\end{align}
Here $\mathcal{S}_\mathrm{BE} (\mathcal{T})$ denotes the set of vacua%
\footnote{The notation BE stands for ``Bethe Equations'', because the 2d F-term equations from the effective twisted superpotential look similar to Bethe ansatz equations.}
of the 3d $\mathcal{N}=2$ theory obtained by extremizing the effective twisted superpotential on $\mathbb{R}^2 \times S^1$ after summing all one-loop contributions from infinitely many massive Kaluza-Klein modes along $S^1$.
The order of the set is equal to the Witten index \cite{Kim:2010mr, Intriligator:2013lca} of the 3d theory
\begin{align}
\bigl| \mathcal{S}_\mathrm{BE}(\mathcal{T}) \bigr|  = \bigl| \textrm{Witten index of $\mathcal{T}$} \bigr| \;.
\end{align}
On the other hand, $\mathcal{H}_\alpha$ is the so-called handle-gluing operator. 

We naturally conjecture that
\begin{align}
\label{3d-3d for Bethe-vacua}
\begin{split}
&\mathcal{S}_\mathrm{BE} (\mathcal{T})  = \frac{\chi_{\rm irred}(M_3;N)}{\mathrm{Hom}[\pi_1(M_3) \rightarrow \mathbb{Z}_N]} \;,
\\
&\mathcal{H}_\alpha = \frac{N}{\bigl| \mathrm{Hom}[\pi_1(M_3) \rightarrow \mathbb{Z}_N] \bigr|} \; \exp \bigl(-2 S^\alpha_1(M_3 ;N) \bigr) \;.
\end{split}
\end{align}
This identification leads to the following generalized 3d-3d relation for twisted indices:
\begin{align}
\begin{split}
&\mathcal{I}_{\Sigma_g} \big( \mathcal{T}_N[M_3] \big) = \sum_{[\mathcal{A}^\alpha] \in \frac{\chi_\mathrm{irred} (M_3;N)}{\mathrm{Hom}[\pi_1(M_3) \rightarrow \mathbb{Z}_N]} }  \left( \frac{N}{\bigl| \mathrm{Hom}[\pi_1(M_3) \rightarrow \mathbb{Z}_N] \bigr|} \; \exp (-2 S_1^{\alpha})   \right)^{g-1}\;
\\
& \textrm{for arbitrary closed hyperbolic 3-manifolds $M_3$} \;.
\label{general 3d-3d for twisted index}
\end{split}
\end{align} 
While in \cite{Gang:2019uay} the 3d-3d relation was proposed for 3-manifolds with vanishing $H_1 (M_3, \mathbb{Z}_N)$, here we generalize it to arbitrary closed hyperbolic 3-manifolds.

\subsection[Integrality of indices at finite $N$]{Integrality of indices at finite \matht{N}}
\label{sec : Integrality check}

To test our proposals, in this subsection we verify a few integrality properties that the indices must satisfy.

\subsubsection{Vanishing of the refined twisted index}
\label{subsec : integrality for refined index}

Since the $R$-charge of $\CT_N[M_3]$ is quantized as in \eqref{quantization of R-charge} and $j_3 \in \frac{1}{2}  \mathbb{Z}$, we expect that 
\begin{equation}
\label{Finiteness of refined index}
\CI_{\rm top} \bigl( q;\CT_N[M_3] \bigr) \,\in\, \mathbb{Z}[q^{1/2},q^{-1/2} ] \;,
\end{equation}
namely that the refined twisted index is a finite Laurent series in $q^{1/2}$ with \emph{integer} coefficients. Indeed, the (refined) twisted index is a Witten index, and it only gets contributions from ground states in the twisted Hilbert space $\mathcal{H}_{\rm top} (S^2)$ which is finite. Since the index is a finite Laurent polynomial in $q^{1/2}$, it converges in the  limit $q=e^\hbar \rightarrow 1$ and we can systematically study it using perturbative methods in complex Chern-Simons theory. 
The 3d-3d relation \eqref{3d-3d for refined index} then translates this rather obvious statement into the following highly non-trivial prediction on the perturbative invariants $\bigl\{ S_n (M_3 ;N) \bigr\}$:
\begin{multline}
\label{prediction from 3d-3d for refined index}
\frac{\bigl| \mathrm{Hom}[\pi_1 (M_3) \rightarrow \mathbb{Z}_N] \bigr|}{N}   \sum_{[\mathcal{A}^\alpha] \in \frac{\chi_{\rm irred} (M_3;N)}{\mathrm{Hom}[\pi_1(M_3) \rightarrow \mathbb{Z}_N]}} \exp \left( 2 \sum_{n=0}^\infty \hbar^{2n} S_{2n+1}^{\alpha} (M_3 ;N) \right)
\\[.5em]
\in  \mathbb{Z}[e^{\hbar/2},e^{-\hbar/2}]  \qquad \textrm{as a formal expansion in $\hbar$.}
\end{multline}
In the following we will check that, in fact, an even stronger statement holds true: the LHS of \eqref{prediction from 3d-3d for refined index} actually \emph{vanishes} for many closed hyperbolic 3-manifolds $M_3$ (at least up to some high power of $\hbar$).

We can offer the following physical argument for such a vanishing. The setup is $N$ M5-branes on $S^2 \times S^1 \times M_3$. The compactification on $S^2$ \cite{Gaiotto:2009we} produces a peculiar class $\mathcal{S}$ theory with empty Coulomb branch but non-trivial Higgs branch. This can be understood by further compactification on $S^1$: the 3d mirror is 3d $\mathcal{N}=4$ $SU(N)$ SYM \cite{Benini:2010uu}, which has hyper-K\"ahler Coulomb branch and empty Higgs branch. However, assuming that the theory $\CT_N[M_3]$ only captures a branch of the moduli space of M5-branes, which in this case corresponds to the Coulomb branch of the class $\CS$ theory \cite{Gukov:2016gkn}, the fact that such a branch is empty implies that the twisted index vanishes.

\paragraph{Example:  \matht{M_3= (S^3 \backslash \mathbf{4}_1)_{P/Q=5}} and \matht{N=2}.} We refer to \eqref{Dehn surgery representation} for our notation of Dehn-surgery representation of closed 3-manifolds. Here $\mathbf{4}_1$ is the figure-eight knot. This is known as Meyerhoff 3-manifold, and its hyperbolic volume is approximately
\begin{align}
\textrm{vol}(M_3) \simeq 0.981369 \;.
\end{align}
The corresponding 3d gauge theory for $N=2$ is \cite{Gang:2017lsr}
\begin{align}
\label{T[M3]-for-M3=(41)-5}
\mathcal{T}_{N=2}[M_3] = \Big( \textrm{$U(1)_{k=-7/2}$ coupled to a single chiral $\Phi$ of charge $+1$} \Big) \;.
\end{align}
The perturbative invariants are:
\begin{align}
& S_0 \left((S^3\backslash \mathbf{4}_1)_5;N=2 \right)  = \textrm{Li}_2 (e^{-Z}) + i \pi (1+2 \ell_z ) Z - \frac{3}2 Z^2 \;,
\nn \\
& S_1  \left((S^3\backslash \mathbf{4}_1)_5;N=2 \right) = \log z - \frac{1}2 \log (3z-4) + \frac{1} 2 \log 2\;, 
\label{Sn for (41)5} \\
&S_2  \left((S^3\backslash \mathbf{4}_1)_5;N=2 \right)=\frac{27 z^3-54 z^2+92 z-64}{24 (3 z-4)^3}\;,
\nn \\
&S_3  \left((S^3\backslash \mathbf{4}_1)_5;N=2 \right)=\frac{5 z \left(21 z^3-49 z^2+36 z-8\right)}{2 (4-3 z)^6}\;,
\nn \\
&S_4  \left((S^3\backslash \mathbf{4}_1)_5;N=2 \right)=\frac{z \left(-19683 z^7+318087 z^6+75762 z^5-2103318 z^4+2881056 z^3 \right)}{720 (3 z-4)^9}\;
\nn \\
& \qquad \qquad  \qquad \qquad \qquad + \frac{z \left(-1056608 z^2-265152 z+169856\right)}{720 (3 z-4)^9}\;,
\nn \\
& S_5 \left((S^3\backslash \mathbf{4}_1)_5;N=2 \right) =\frac{z \left(-11907 z^9+671895 z^8-1057914 z^7-2768940 z^6+8509962 z^5 \right)}{24 (4-3 z)^{12}}\;
\nn \\
& \qquad \qquad  \qquad  \qquad \qquad  + \frac{z \left(-7816248 z^4+2119680 z^3+809536 z^2-511872 z+55808\right)}{24 (4-3 z)^{12}}\;.
\nn
\end{align}
Here $z:=e^{Z}$ and $\ell_z \in \mathbb{Z}$ take different values on each irreducible flat connection. 
There are 4 irreducible $SL(2,\mathbb{C})$ flat connections on $M_3$, which correspond to the following 4 Bethe vacua:
\begin{align}
\exp \big( \partial_{Z} S_0 \big)
= 1 \qquad\Rightarrow\qquad  \mathcal{S}_\mathrm{BE} \Bigl(\mathcal{T}_{N=2}\bigl[ M_3 = (S^3\backslash \mathbf{4}_1)_5 \bigr] \Bigr) = \left\{ z \;\middle|\; \frac{1-z}{z^4}=1 \right\} \;. 
\end{align}
The manifold $M_3$ has vanishing $H_1 (M_3, \mathbb{Z}_2)$ and there are no $\mathbb{Z}_2$ flat connections. This is compatibile with the Witten index computation \cite{Intriligator:2013lca}:
\begin{align}
&\textrm{Witten index of $\big( U(1)_k  + \textrm{fundamental }\Phi \big) = |k|+1/2$} \label{Witten index}
\\
&\Rightarrow\;\; \textrm{Witten index of $\mathcal{T}_{N=2}\bigl[ M_3 = (S^3\backslash \mathbf{4}_1)_5 \bigr]$ = 4}\;. \nonumber
\end{align}
For each solution to the Bethe vacua equation, there exists a solution $Z$ satisfying $\partial_{Z} S_0 = 0$ for a proper choice of $\ell_z \in \mathbb{Z}$: this fixes $\ell_z$. Except for the classical part $S_0$, the perturbative invariants are independent of the choice of $\ell_z$.
The numerical values of $Z, \ell_z$ on the Bethe vacua and their classical actions are
\begin{equation}
\label{4 Bethe-vacua}
\begin{aligned}
\mathcal{A}^{\overline{\rm geom}} \;:\; Z &=  0.061412\, +1.33528 i \;  & (\ell_z =0) ,\quad S_0 &= -1.52067-0.981369 i\;,
\\
\mathcal{A}^{\rm geom} \;:\; Z &=  0.061412\, -1.33528 i \; & (\ell_z =-1) ,\quad S_0 &= -1.52067+0.981369 i\;,
\\
\mathcal{A}^{\rm 3rd} \;:\; Z &=  0.199461\, + \pi i \; & (\ell_z =1) ,\quad S_0 &= -15.5579\;,
\\
\mathcal{A}^{\rm 4th} \;:\; Z &= -0.322285\; & (\ell_z =-1) ,\quad S_0 &= 2.14992\;.
\end{aligned}
\end{equation}
Note that $\mathrm{Im}\bigl[ S_0^{\rm geom} \bigr] = - \mathrm{Im}\bigl[ S_0^{\overline{\rm geom}} \bigr] = \textrm{vol}(M_3)$. The $\mathcal{A}^{\rm 3rd}$ and $\mathcal{A}^{\rm 4th}$ are irreducible flat connections in $SU(2) \subset SL(2,\mathbb{C})$ and have vanishing $\textrm{Im}[S_0]$. Using the perturbative invariants (recall $q=e^\hbar$), one can check that
\begin{align}
&\mathcal{I}_\mathrm{top} \Bigl( q;\, \mathcal{T}_{N=2}\bigl[ M_3 = (S^3\backslash \mathbf{4}_1)_5 \bigr] \Bigr) 
\nn \\
&=\frac{1}{2} \sum_{\frac{1-z}{z^4}=1} \exp \bigg{(}2 S_1(M_3; N) + 2 S_3(M_3; N)  \hbar^2 + 2 S_5(M_3; N)  \hbar^4   + \CO(\hbar^6) \bigg{)}\bigg{|}^{N=2}_{M_3 = (S^3\backslash \mathbf{4}_1)_5}
\nn \\
&= 0 + \CO (\hbar^6) \;. 
\end{align}
Surprisingly, the refined twisted index vanishes (at least up to a certain power of $\hbar$) in a highly non-trivial way. We have checked that the vanishing persists up to $\CO(\hbar^{10})$.

\paragraph{Example: \matht{M_3= \textrm{(Weeks manifold)}:= (S^3 \backslash \mathbf{5}^2_1 )_{P_1/Q_1 =- 5,\, P_2/Q_2 = -5/2}}.} Here $\mathbf{5}_1^2$ is the Whitehead link. This manifold is the smallest orientable closed hyperbolic 3-manifold \cite{gabai2009minimum}, and its volume is
\begin{align}
\textrm{vol}(M_3) \simeq 0.942707 \;. 
\end{align}
The corresponding 3d gauge theory for $N=2$ is \cite{Gang:2017lsr}
\begin{align}
\mathcal{T}_{N=2}[M_3] = \Big( \textrm{$U(1)_{k=-5/2}$ coupled to a single chiral $\Phi$ of charge $+1$} \Big) \;.
\label{3d-theory-for-weeks}
\end{align}
The Witten index  \eqref{Witten index} for the theory is $3$ and there are 3 irreducible flat connections given by the following Bethe equation:
\begin{align}
\exp \big(\partial_Z S_0 \big) = 1 \qquad \Rightarrow \qquad \frac{z-1}{z^3}=1 \;.
\end{align}
The solutions and the corresponding classical actions are
\begin{equation}
\begin{aligned}
\mathcal{A}^{\overline{\rm geom}} \;:\; Z &=  0.1406\, + 0.703858 i \;  & (\ell_z =0) ,\quad S_0 &= 1.1852 - 0.942707 i\;,
\\
\mathcal{A}^{\rm geom} \;:\; Z &=  0.1406\, - 0.703858 i \; & (\ell_z =0) ,\quad S_0 &= 1.1852 + 0.942707 i\;,
\\
\mathcal{A}^{\rm 3rd} \;:\; Z &=  0.2812\, + \pi i \; & (\ell_z =1) ,\quad S_0 &= -10.5951 \;.
\end{aligned}
\end{equation}
The manifold has vanishing $H_1 (M_3, \mathbb{Z}_2)$ and there are no $\mathbb{Z}_2$ flat connections. The perturbative invariants $\{S_n\}$ up  to $n=5$  for the irreducible flat connections are
\begin{align}
& S_0(\textbf{Weeks};N=2) = \mathrm{Li}_2 (e^{-Z}) + 2\pi i \ell_z Z - Z^2 \;,
\nn \\
&S_1 (\textbf{Weeks};N=2)  =-\frac{1}{2} \log \left(1-\frac{3}{2 z}\right)\;,
\label{perturbative invariants Weeks} \\
&S_2 (\textbf{Weeks};N=2)  = \frac{46 z^2-72 z+27}{24 (2 z-3)^3}\;,
\nn \\
&S_3(\textbf{Weeks};N=2) = \frac{z \left(5 z^4+9 z^3-41 z^2+36 z-9\right)}{2 (3-2 z)^6}\;,
\nn \\
&S_4  (\textbf{Weeks};N=2) = \frac{(z-1) z \left(988 z^6+45700 z^5-9642 z^4-149328 z^3+140103 z^2\right)}{720 (2 z-3)^9}\;,
\nn \\
&\qquad \qquad \qquad \qquad \quad + \frac{(z-1) z \left(-18225 z-9477\right)}{720 (2 z-3)^9}\;,
\nn % \\
\end{align}
\begin{align}
&S_5 (\textbf{Weeks};N=2) = \frac{(1-z) z \left(4 z^9-4022 z^8-34158 z^7+64404 z^6+74400 z^5-209358 z^4\right)}{24 (3-2 z)^{12}}\;
\nn \\
&\qquad \qquad \qquad \qquad \quad  + \frac{(1-z) z \left(121041 z^3+1296 z^2-15795 z+2187\right)}{24 (3-2 z)^{12}}\;.
\nn 
\end{align}
Using the perturbative invariants, one can check that
\begin{align}
&\mathcal{I}_\textrm{top} \Bigl(q; \mathcal{T}_{N=2}[M_3 =\textrm{\bf Weeks}] \Bigr) 
\nn \\
&=\frac{1}{2} \sum_{\frac{z-1}{z^3}=1} \exp \bigg{(}2 S_1(M_3; N) + 2 S_3(M_3; N)  \hbar^2 + 2 S_5(M_3; N)  \hbar^4   + \CO(\hbar^6) \bigg{)}\bigg{|}^{N=2}_{M_3 = \textrm{\bf Weeks}}
\nn \\
&= 0 + \CO(\hbar^6)\;. 
\end{align}
Again, the refined index vanishes in a highly non-trivial way --- we have checked the vanishing up to  $\CO(\hbar^{10})$.

\paragraph{Example: \matht{M_3=  (S^3 \backslash \mathbf{5}^2_1 )_{P_1/Q_1 =-6 ,\, P_2/Q_2 = -3/2}}.}
This manifold is the third smallest orientable closed hyperbolic 3-manifold \cite{gabai2009minimum}, and its volume is
\begin{align}
\textrm{vol}(M_3) \simeq 1.014942  \;. 
\end{align}
The corresponding 3d gauge theory for $N=2$ is \cite{Gang:2017lsr}
\begin{align}
\mathcal{T}_{N=2}[M_3] = \Bigl( \textrm{$U(1)_{k=-3/2}$ coupled to a single chiral $\Phi$ of charge $+1$} \Bigr) \;.
\label{3d-theory-for-3rd}
\end{align}
The classical Chern-Simons action is
\begin{equation}
S_0 = \mathrm{Li}_2(e^{-Z}) + 2\pi i \ell_z Z - \frac12 Z^2 \;.
\end{equation}
The Witten index of (\ref{3d-theory-for-3rd}) is $2$, and the two solutions to the Bethe equations $\exp( \partial_Z S_0)=1$ correspond to the equivalence classes of the two geometric flat connections (in this case $H_1(M_3, \BZ_2) = \BZ_2$).
Using the state-integral model for the 3-manifold, we have checked that the refined index computed using the 3d-3d relation  \eqref{3d-3d for refined index}   vanishes up to $\CO(\hbar^{10})$.

\subsubsection{Integer expansion of the twisted index}

On a generic Riemann surface $\Sigma_g$, thus with no $q$-refinement, we expect
\begin{equation}
\mathcal{I}_{\Sigma_g} \bigl( \mathcal{T}_{N=2}[M_3] \bigr) \in \mathbb{Z} \;.
\end{equation}
According to  the 3d-3d relation \eqref{general 3d-3d for twisted index}, the (unrefined) twisted index is determined by the one-loop invariants $S^\alpha_1 (M_3 ;N)$. The one-loop part is simply related to the mathematical quantity called analytic Ray-Singer torsion, see \eqref{Classical/1-loop}. Computing such topological quantity using its definition \eqref{Def of Ray-Singer torsion} is quite a challenging task, since we need to know the full spectrum of the Laplacians on the 3-manifold. There are two simpler alternative ways to compute the Ray-Singer torsion: i) using a state-integral model for complex Chern-Simons theory, and ii) using Cheeger-Muller theorem. See appendix~\ref{App: perturbative invariants} for a brief introduction to the state-integral model. Cheeger-Muller theorem claims equivalence between the analytic Ray-Singer torsion and the combinatorial Reidermeister torsion. The combinatorial torsion can be computed from so-called Fox calculus on the $SL(N,\mathbb{C})$ representations of $\pi_1 (M_3)$. In \cite{Dimofte:2012qj} it is explicitly checked that the two simpler approaches give the same answer in a large number of examples. In this subsection, we give the explicit expression for the combinatorial Reidermeister torsion $\mathrm{Tor}[\CA^\alpha;M_3]$ (equal to the analytic Ray-Singer torsion) twisted by $SL(2,\mathbb{C})$ irreducible flat connections $\CA^\a$ on $M_3 =(S^3\backslash \mathbf{4}_1)_{P/Q}$ and check the integrality of $\mathcal{I}_{\Sigma_g} \bigl( \mathcal{T}_{N=2}[M_3] \bigr)$  in \eqref{general 3d-3d for twisted index}.

\paragraph{Example: \matht{(S^3\backslash \mathbf{4}_1)_{P/Q}}.}
These manifolds are hyperbolic for all  $P/Q \in \mathbb{Q} \cup \{\infty \}$ but the following 10 exceptional slopes:
\begin{align}
\textrm{Exceptional slopes of $S^3\backslash \mathbf{4}_1$ \;:\; $P/Q =\{0, \pm 1, \pm 2, \pm 3, \pm 4, \infty\}$}\;.
\end{align}
The fundamental groups of these 3-manifolds are
\begin{align}
\label{fundamental groups}
\begin{split}
&\pi_1 (S^3 \backslash \mathbf{4}_1)  = \bigl\langle a,b : a b^{-1} a^{-1} b a = b a b^{-1} a^{-1} b  \bigr\rangle\;,
\\
&\pi_1 \left(M_3  \right) = \bigl\langle a,b :  a b^{-1} a^{-1} b a = b a b^{-1} a^{-1} b \,,\;  \mathbf{m}^P  \mathbf{l}^Q=1 \bigr\rangle  \;,\; \textrm{where}
\\
&\mathbf{m} := a, \quad \mathbf{l} :=  a b^{-1} a b a^{-2} b a b^{-1}a^{-1}\;. 
\end{split}
\end{align}
The set $\chi_{\rm irred}\big(M_3= (S^3\backslash \mathbf{4}_1)_{P/Q} ; N=2 \big)$ can be obtained by solving the following matrix equations:
\begin{align}
\label{chi-41-PQ}
\chi_{\rm irred}\bigl(M_3;N=2\bigr) = \left\{ A = \begin{pmatrix} 
m & 1 \\
0 & m^{-1} 
\end{pmatrix},\; B  = \begin{pmatrix} 
m & 0 \\
y & m^{-1} 
\end{pmatrix} : \textrm{matrix eqns.} \right\} / \mathbb{Z}_2\;,
\end{align}
where the matrix equations are
\begin{align}
\begin{split}
&AB^{-1}A^{-1}BA=BAB^{-1}A^{-1}B \qquad\textrm{and}\qquad M^P L^Q = \mathbb{I}  \qquad\textrm{with}
\\
&M:=A\;, \;\; L = AB^{-1}ABA^{-2} BAB^{-1}A^{-1}\;.
\end{split}
\end{align}
Indeed, one can show%
\footnote{From the first line of (\ref{fundamental groups}) it follows $b = (a^{-1} b a b^{-1}) a (b a^{-1} b^{-1}a)$.}
that $a$ and $b$ are related to each other by a conjugation in $\pi_1 (M_3)$, and so their corresponding  holonomy matrices $A$ and $B$ are related by a conjugation in $SL(2,\mathbb{C})$. Using the $SL(2,\mathbb{C})$ conjugation, we fix the matrices $A$ and $B$ as above. There is a residual $\mathbb{Z}_2$  gauge symmetry acting as
\begin{align}
\mathbb{Z}_2 \;:\; (m,y) \leftrightarrow (m^{-1},y)\;.
\end{align}
The matrix equations give  algebraic equations for $m$ and $y$, which can be solved numerically. When $P/Q=5$, we have the following 4 solutions which can be identified with the four Bethe vacua in \eqref{4 Bethe-vacua}:
\begin{equation}
\begin{aligned}
(m,y) &= (1.20331 +0.780836 i \,,\; 0.992448 +0.513116 i) && \leftrightarrow\quad \mathcal{A}^{\overline{\rm geom}}
\\
(m,y) &= (1.20331 -0.780836 i \,,\; 0.992448-0.513116 i) && \leftrightarrow\quad \mathcal{A}^{\rm geom}
\\
(m,y) &= (0.164478 +0.986381 i \,,\; 3.49022)\;  && \leftrightarrow\quad \mathcal{A}^{\rm 3rd}
\\
(m,y) &= (-0.452583+0.891722 i \,,\; 2.52489)\; && \leftrightarrow\quad \mathcal{A}^{\rm 4th} \;.
\end{aligned}
\end{equation}
The torsion can be computed from the following formula (see the appendix of \cite{Gang:2019uay} for the derivation):
\begin{align}
\label{torsion 41-PQ}
\begin{split}
&\exp\bigl( -2S^\alpha_1 (M_3; N=2) \bigr)\Big{|}_{M_3 = (S^3\backslash \mathbf{4}_1)_{P/Q}}
\\
&= \left( P \, \Bigl( \ell - \frac{1}\ell \Bigr) \, \frac{m^4}{(m^4-1)(4-2m^2+4m^4)} +Q \right) \frac{ 2 m^2 + \frac{2}{m^2}-1 }{ (1-m^{2R} \ell^{2S}) (1-m^{-2R} \ell^{-2S}) } \;.
\end{split}
\end{align}
Here $\ell$ is defined from the following relation:
\begin{align}
L = \begin{pmatrix} 
\ell & * \\
0 & \ell^{-1} 
\end{pmatrix} \;.
\end{align}
Indeed, after solving the matrix equations, the two $SL(2,\mathbb{C})$ matrices $M$ and $L$ commute since $\mathbf{m}\mathbf{l} = \mathbf{l}\mathbf{m}$ in $\pi_1 (S^3\backslash \mathbf{4}_1)$ and in $\pi_1(M_3)$, as one can show.%
\footnote{From the first line of (\ref{fundamental groups}) follow $a^2 b^{-1} ab a^{-1} = ab^{-1}aba^{-1}b$ and $b^{-1}a^{-1}bab^{-1} = a^{-1}bab^{-1}a^{-1}$.}
So, after solving the matrix equations, $L$  takes the upper-triangular form as above. One can show that the constraint of $\pi_1(S^3\backslash\mathbf{4}_1)$ implies
\begin{align}
2 + \ell + \ell^{-1} - m^4 + m^2 + m^{-2} - m^{-4} = 0 \;,
\end{align}
which is the A-polynomial of $\mathbf{4}_1$. Then, $(R,S)$ are coprime integers chosen such that $PS-QR =1$. For given $(P,Q)$, the choice is not unique  but  the above torsion  is independent of the choice since the matrix equation $M^P L^Q= \mathbb{I}$ implies $m^P \ell^Q=1$.
By applying the formula to $P/Q=5$, we find that
\begin{align}
\begin{split}
&\exp\bigl(-2S_1^\alpha (M_3; N=2)\bigr)\big{|}^{\alpha = \overline{\rm geom}}_{M_3 = (S^3\backslash \mathbf{4}_1)_{P/Q=5}}= 1.90538 -0.568995 i\;,
\\
&\exp\bigl(-2S_1^\alpha (M_3; N=2)\bigr)\big{|}^{\alpha = {\rm geom}}_{M_3 = (S^3\backslash \mathbf{4}_1)_{P/Q=5}}= 1.90538 +0.568995 i\;,
\\
&\exp\bigl(-2S_1^\alpha (M_3; N=2)\bigr)\big{|}^{\alpha = {\rm 3rd}}_{M_3 = (S^3\backslash \mathbf{4}_1)_{P/Q=5}}= -2.57085\;,
\\
&\exp\bigl(-2S_1^\alpha (M_3; N=2)\bigr)\big{|}^{\alpha = {\rm 4th}}_{M_3 = (S^3\backslash \mathbf{4}_1)_{P/Q=5}}= -1.73992\;. \label{numerical torsion for 41-5}
\end{split}
\end{align}
One can check  that the numerical values are equal to $e^{-2S_1^\alpha}$ in \eqref{Sn for (41)5} for the Bethe vacua \eqref{4 Bethe-vacua} computed from the state-integral model.

Combining the torsion computations in \eqref{chi-41-PQ} and \eqref{torsion 41-PQ} with the 3d-3d relation in \eqref{general 3d-3d for twisted index}, we obtain the following concrete expression for the twisted index:
\begin{multline}
\mathcal{I}_{\Sigma_g} \big( \mathcal{T}_{N=2}[M_3] \big) = \frac{1}{\bigl| \mathrm{Hom}[\pi_1(M_3)\rightarrow \mathbb{Z}_2] \bigr|} 
\\
\times   \sum_{(m,y) \in \chi_{\rm irred} \textrm{ in } \eqref{chi-41-PQ}} \left( \frac{2}{\bigl| \mathrm{Hom}[\pi_1(M_3)\rightarrow \mathbb{Z}_2] \bigr|} \; \Bigl( \exp(-2S_1)\textrm{ in \eqref{torsion 41-PQ}} \Bigr) \right)^{g-1}
\end{multline}
for $M_3 = (S^3\backslash \mathbf{4}_1 )_{P/Q}$.
Note that
\begin{align}
\bigl| \mathrm{Hom}[\pi_1 (M_3) \rightarrow \mathbb{Z}_2] \bigr|= \begin{cases}
1 \quad  \textrm{for odd $P$}\;, \\
2 \quad \textrm{for even $P$}\;.\\
\end{cases} 
\end{align}
Using the expression above, one can check that $\mathcal{I}_{\Sigma_g} \bigl( \mathcal{I}_{N=2}[M_3] \bigr)  \in \mathbb{Z}$, which is quite non-trivial.  

For example, using the torsions in \eqref{numerical torsion for 41-5} for the case $P/Q=5$, one obtains
\begin{align}
\label{integers TTindex 1st example}
\begin{split}
&\mathcal{I}_{\Sigma_g} \bigl( \mathcal{T}_{N=2}[M_3] \bigr)\bigg{|}_{M_3 = (S^3\backslash \mathbf{4_1})_{P/Q=5}} 
\\
&=\sum_\alpha  2^{g-1}\exp\Bigl( -2(g-1)\, S^\alpha_1 (M_3; N=2) \Bigr)  \bigg{|}_{M_3 = (S^3\backslash \mathbf{4_1})_{P/Q=5}} 
\\
&= \big{\{} 0_{g=0},\; 4_{g=1},\; -1_{g=2},\; 65_{g=3},\;-97_{g=4},\; \ldots \big{\}} \;.
\end{split}
\end{align}
In fact, we can give a precise algebraic characterization of these numbers. The Bethe vacua \eqref{4 Bethe-vacua} are the four solutions to the algebraic equation $z^4 + z - 1 = 0$.
From \eqref{Sn for (41)5}, the Ray-Singer torsion contribution from each irreducible flat connection is
\be
2 \, e^{-2 S_1} = \frac{3z-4}{z^2} \equiv w \;.
\ee
One can easily show that the Bethe ansatz algebraic equation implies
\be
w^4 + w^3 - 32 w^2 + 283 = 0 \;.
\ee
In other words, the four values of $w$ are the roots of an algebraic equation with integer coefficients. This implies that $\sum_{i=1}^4 w_i^{g-1} \in \mathbb{Z}$ for all $g \geq 1$, and it allows to determine the integers values of the index in (\ref{integers TTindex 1st example}) in an algebraic way from the coefficients of the equation. Since the equation has no linear term, it also implies (for $g=0$) that $\sum_{i=1}^4 w_i^{-1}=0$ which is our vanishing conjecture.

The case of Weeks' manifold can be treated in a similar way. From (\ref{perturbative invariants Weeks}), the Bethe ansatz equation --- or equation for the irreducible flat connections --- is $z^3 - z+1=0$. The contributions from Ray-Singer torsion are $2\, e^{-2S_1} = (2z-3)/z \equiv w$,
and one easily shows that
\be
w^3 - 3w^2 - 23 = 0 \;.
\ee
The fact that $w$'s satisfy an algebraic equation with integer coefficients and no linear term implies that $\sum_{i=1}^3 w_i^{g-1} \in \BZ$ for $g \geq 1$, and that it vanishes for $g=0$.

In the case of $(S^3\setminus \mathbf{4}_1)_{P/Q}$ with $P/Q=6$, there are 6 irreducible flat connections in $\chi_\textrm{irred}(M_3; N=2)$ but there are only 3 equivalence classes  $[\mathcal{A}^\alpha] \in \frac{\chi_\textrm{irred}}{\mathrm{Hom}[\pi_1 (M_3) \rightarrow \mathbb{Z}_2]}$. The numerical values of the torsion for the three classes are 
\begin{multline}
\Bigl\{ \exp\bigl( -2S_1^\alpha (M_3; N=2) \bigr)  \Bigr\}_{[\mathcal{A}^\alpha] \in \frac{\chi_\textrm{irred}}{\mathrm{Hom}[\pi_1 (M_3) \rightarrow \mathbb{Z}_2]} } \qquad \textrm{for $M_3 = (S^3\backslash \mathbf{4}_1)_{P/Q=6}$}
\\
= \bigl\{ 4.30835\, -1.99637 i \,,\; 4.30835\, +1.99637 i \,,\; -2.61671 \bigr\} \;.
\end{multline}
Then, we see that
\begin{align}
\begin{split}
&\mathcal{I}_{\Sigma_g} \bigl( \mathcal{T}_{N=2}[M_3] \bigr)\bigg{|}_{M_3 = (S^3\backslash \mathbf{4_1})_{P/Q=6}} 
\\
&\sum_{[\mathcal{A}^\alpha] \in \frac{\chi_\textrm{irred}}{\mathrm{Hom}[\pi_1 (M_3) \rightarrow \mathbb{Z}_2]} }  \exp\Bigl( -2(g-1) \, S^\alpha_1 (M_3; N=2) \Bigr)  \bigg{|}_{M_3 = (S^3\backslash \mathbf{4_1})_{P/Q=6}} 
\\
&= \big{\{} 0_{g=0},\; 3_{g=1},\; 6_{g=2},\; 36_{g=3},\;39_{g=4},\; \ldots \big{\}}\;.
\end{split}
\end{align}
This integrality constitutes a non-trivial consistency check for all the subtle factors in the 3d-3d relation \eqref{general 3d-3d for twisted index}. These factors will play an important role in the computation of the $\log N$ subleading corrections to the twisted indices at large $N$.

%%%%%%%%%%%%%%%%%%%%%%%%%%%%%%%%%%%%%%%%%%%%%%%%%%%%%%%%%%%%%%%%

\section{Large $\boldsymbol{N}$ limit of the indices}
\label{Sec:LargeN}

We consider now the large $N$ limit of the three types of indices of $\mathcal{T}_N[M_3]$ using the 3d-3d relations in \eqref{3d-3d for SCI}, \eqref{3d-3d for refined index} and \eqref{general 3d-3d for twisted index}. Through the 3d-3d relations, all indices are related to the holomorphic blocks \eqref{CS holomprhic block}
of $SL(N,\mathbb{C})$ Chern-Simons  theory on $M_3$. The blocks can be perturbatively expanded in the holomorphic coupling.  We have two expansion parameters,  $1/N$ and $\hbar$. We first consider the perturbative expansion in $\hbar$, and then take the large $N$ limit of the perturbative expansion coefficients $\bigl\{ S_n(M_3;N) \bigr\}$.

\subsection{Perturbative Chern-Simons invariants}

There are two canonical $SL(N,\mathbb{C})$ irreducible flat connections on any hyperbolic 3-manifold $M_3$, denoted $\CA^{\rm geom}_N$ and $\CA_N^{\overline{\rm  geom}}$:
\begin{align}
\CA^{\rm geom}_N =\rho_N (\omega + i e)\;, \quad \CA^{\overline{\rm geom}}_N =\rho_N (\omega - i e)\;.
\end{align}
According to Mostow's rigidity theorem \cite{mostow1968quasi}, there is a unique hyperbolic metric on $M_3$ satisfying $R_{\mu\nu}=-2 g_{\mu\nu}$.   Here $e$ and $\omega$ are dreibein and spin connection of the hyperbolic structure, respectively. Both of them can be considered as $\mathfrak{so}(3)$-valued 1-forms and they form two $SL(2,\mathbb{C})$ flat connections, $\omega \pm i e$. On the other hand, $\rho_N : SL(2) \rightarrow SL(N)$ is the principal embedding. A characteristic property of the two irreducible flat connections is 
\begin{equation}
\label{character property of Ageom}
\mathrm{Im} [S_0^{\overline{\rm geom}}] \leq   \textrm{Im} [S_0^\alpha]  \leq \textrm{Im} [S_0^{\rm geom}] \qquad\qquad
\forall \;\; [\mathcal{A}^\alpha] \in \frac{\chi_{\rm irred}(M_3;N)}{\mathrm{Hom}[\pi_1 (M_3)\rightarrow \mathbb{Z}_N]} \;.
\end{equation}
The equality holds for and only for $\alpha = {\rm geom}$ or $\alpha = \overline{\rm geom}$. 
The classical and one-loop parts around the flat connections can be  expressed in terms of the hyperbolic volume of the 3-manifold, $\mathrm{vol}(M_3)$, and the complex length spectrum $\bigl\{ \ell_{\mathbb{C}} (\gamma) \bigr\}$:
\begin{align}
\label{large N of S0, S1}
\textrm{Im}\big[ S^{\rm geom}_0(M_3;N) \big] &= \frac{(N^3-N)}6 \, \textrm{vol}(M_3) \;,
\\
\textrm{Re} \big[S^{\rm geom}_1(M_3;N) \big] &= -\frac{\textrm{vol}(M_3)}{12\pi } \bigl( 2N^3-N-1 \bigr)-\frac{1}2\sum_{\gamma}\sum_{m=1}^{N-1} \sum_{k=m+1}^{\infty}  \log \bigl| 1-e^{-k \, \ell_{\mathbb C}(\gamma)} \bigr| \;.
\nn
\end{align}
Here we defined
\begin{align}
\textrm{vol}(M_3) := \big( \textrm{volume of $M_3$ measured in the unique hyperbolic metric} \big) \;.
\end{align}
Moreover $\sum_{\gamma}$ is the sum over the non-trivial primitive geodesics $\gamma \in M_3$, and $\ell_{\mathbb{C}}(\gamma)$ is the complexified geodesic length of $\gamma$, which is defined by the following relation:
\begin{align}
\mathrm{Tr} \; \mathrm{Pexp} \left(-\oint_{\gamma} \CA_{N=2}^{\rm geom} \right)= e^{\frac{1}2 \ell_{\mathbb{C}}(\gamma)}+ e^{-\frac{1}2 \ell_{\mathbb{C}}(\gamma)}\;, \quad \mathrm{Re} [\ell_{\mathbb{C}}]>0\;.
\end{align}
The real part of $\ell_{\mathbb{C}}$ measures the geodesic length.

Let us explain the origin of the two formulas in (\ref{large N of S0, S1}). The classical part of the Chern-Simons action can be  obtained using the following computation ($h_1$ and $h_2$ are non-zero elements in $\mathfrak{sl}(2, \mathbb{C})$):
\begin{align}
\label{large N of S0, S1-2}
&\textrm{Im}\bigl[ S_0^{\rm geom}(M_3;N) \bigr]= \textrm{Im}\Bigl[ -2\pi \, \mathrm{CS}[\mathcal{A}^{\rm geom}_N, M_3] \Bigr]
\\
& = \frac{\textrm{Tr} \bigl( \rho_N (h_1) \, \rho_N (h_2) \bigr)}{\textrm{Tr}\bigl( \rho_{N=2} (h_1) \, \rho_{N=2} (h_2)\bigr)} \; \textrm{Im} \Bigl[  -2\pi \, \mathrm{CS}[\mathcal{A}^{\rm geom}_{N=2}, M_3] \Bigr]
\nn \\
&=\frac{(N^3-N)}{6} \; \textrm{Im} \Bigl[ -2\pi \, \mathrm{CS}[\mathcal{A} = \omega+ i e, M_3] \Bigr] = -\frac{(N^3-N)}{24}   \int_{M_3 }\sqrt{g}(R +2)\big{|}_{R_{\mu \nu} = - 2g_{\mu\nu}}
\nn \\
&= \frac{(N^3-N)}{6}   \int_{M_3 }\sqrt{g}  = \frac{(N^3-N)}{6} \; \textrm{vol}(M_3)\;.
\nn
\end{align}
The expression for the one-loop part $S_1^{\rm geom}$ is derived in \cite{Gang:2019uay} from results \cite{muller2012asymptotics, park2019reidemeister}  in mathematics using Selberg's trace formula. To see the $1/N$ expansion more explicitly, we can write 
\begin{align}
\label{Selberg's expansion}
\begin{split}
&\sum_{\gamma}\sum_{m=1}^{N-1} \sum_{k=m+1}^{\infty}  \log \bigl| 1-e^{-k \, \ell_{\mathbb C}(\gamma)} \bigr| 
\\
&= -\mathrm{Re} \sum_\gamma \sum_{s=1}^{\infty}\frac{1}s \left(\frac{e^{-s \ell_{\mathbb{C}}}}{1-e^{-s  \ell_{\mathbb{C}}}}\right)^2 + \mathrm{Re} \sum_\gamma \sum_{s=1}^{\infty}\frac{1}s \left(\frac{e^{-s(N+1)\ell_{\mathbb{C}}/2 }}{1-e^{-s  \ell_{\mathbb{C}}}}\right)^2\;.
\end{split}
\end{align} 
The first term is independent of $N$ while the second one is exponentially suppressed at large $N$.

The higher perturbative invariants $S_n^{\rm geom}$ ($n\geq 2$) are conjectured to behave at large $N$ as \cite{Gang:2014ema}\footnote{In \cite{Gang:2014ema}, $\CA^{\overline{\rm geom}}$ is denoted as $\CA^{\rm conj}$.}
\begin{align}
\begin{split}
\textrm{Im} \bigl[ S^{\rm geom}_2(M_3;N) \bigr] &= - \frac{N^3}{24\pi^2} \, \mathrm{vol}(M_3) \, + \, \textrm{subleading} \;,
\\
\lim_{N\rightarrow \infty} \; \frac{1}{N^3} \; \mathrm{Re} \bigl[ S^{\rm geom}_{2n+1}(M_3;N) \bigr] &= 0  \qquad (n\geq 1)\;,
\\
\lim_{N\rightarrow \infty}\;  \frac{1}{N^3} \; \mathrm{Im} \bigl[ S^{\rm geom}_{2n}(M_3;N) \bigr] &= 0   \qquad (n\geq 2)\;.
\label{large N of Sn n>=2}
\end{split}
\end{align}
Unlike $S^{\rm geom}_0$ and $S^{\rm geom}_1$, we have conjectures only for the leading $\CO(N^3)$ terms at large $N$.  
The large $N$ behavior of  $S^{\overline{\rm geom}}$ can be obtained using the following general fact:
\begin{align}
S^{\overline{\a}}_n  = (S^{\a}_n)^*\;.
\end{align}
The conjecture \eqref{large N of Sn n>=2} has been numerically tested in many examples up to seven loops.
One way to understand the conjecture is the S-duality of complex Chern-Simons theory \cite{Dimofte:2011jd} relating $\hbar \leftrightarrow - \frac{4\pi^2}\hbar$.  At finite $N$, the S-duality is only seen  after Borel-Pad\'e resummation \cite{Gang:2017hbs}. At large $N$, on the other hand, the perturbative expansion becomes a convergent series as far as the leading $\CO(N^3)$ behavior is concerned, and one can see the S-duality directly in the $\CO(N^3)$ coefficients of the perturbative invariants.

\subsection{Superconformal index}

\subsubsection[Leading $N^3$ behavior]{Leading \matht{N^3} behavior}

Let us set $q:=e^{-\omega}$ with $\mathrm{Im}\;\omega<0$ and consider an expansion for $|\omega|\ll 1$. Combining the 3d-3d relation \eqref{3d-3d for SCI} with the large $N$ behavior \eqref{large N of S0, S1} and \eqref{large N of Sn n>=2}, we find
\begin{align}
&\CI_{\rm sci} \bigl( q;\CT_{N}[M_3] \bigr)  =     \frac{\bigl| \mathrm{Hom}[\pi_1 (M_3) \rightarrow \mathbb{Z}_N] \bigr|}{N}   \sum_{[\mathcal{A}^\alpha] \in \frac{\chi_{\rm irred} (M_3;N)}{\textrm{Hom}[\pi_1 (M_3) \rightarrow \mathbb{Z}_N]} } B^\alpha (q) \; B^{\bar\alpha}(q^{-1}) 
\nn \\
&\xrightarrow{N\rightarrow \infty} \quad \frac{\bigl| \mathrm{Hom}[\pi_1 (M_3) \rightarrow \mathbb{Z}_N] \bigr|}{N}  \;  B^{\rm geom} (q)  \, B^{\overline{\rm geom}}(q^{-1})   + \ldots
\\
&=   \frac{\bigl| \mathrm{Hom}[\pi_1 (M_3) \rightarrow \mathbb{Z}_N] \bigr|}{N}  \;  \exp \left(\sum_{n=0}^\infty (-\omega)^{n-1} S_{n}^{\rm geom} \right)   \exp \left(\sum_{n=0}^\infty \omega^{n-1} (S_{n}^{\rm geom})^* \right)    + \ldots
\nn \\
&=    \frac{\bigl| \mathrm{Hom}[\pi_1 (M_3) \rightarrow \mathbb{Z}_N] \bigr|}{N}  \;  \exp \left\{ \biggl( - \frac{ i N^3}{3\omega} - \frac{N^3}{3\pi} + \frac{i\omega N^3}{12\pi^2} \biggr) \mathrm{vol}(M_3) + \textrm{subleading in $\tfrac1N$} \right\} + \ldots
\nn
\end{align}
where dots stand for exponentially smaller terms in the $1/N$ expansion. In particular, because of (\ref{character property of Ageom}), all irreducible flat connections $\CA^\alpha$ other than $\alpha=\textrm{geom}$ give an exponentially suppressed contribution to the second line.
We thus find
\begin{align}
\label{final result SCI}
\begin{split}
\log \CI_{\rm sci} \bigl( q;\CT_N[M_3] \bigr) &= \frac{iN^3}{12\pi^2} \, \mathrm{vol}(M_3) \, \frac{( \omega + 2\pi i )^2}\omega + \bigl( \textrm{subleading in $1/N$} \bigr)
\\
& = \frac{ i L^2}{8  G_{(4)}} \, \frac{(\omega + 2\pi i )^2}\omega + \bigl( \textrm{subleading in $G_{(4)}/L^2$ } \bigr) \;.
\end{split}
\end{align}
Here $L$ is the radius of AdS$_4$ and $G_{(4)}$ is the 4d Newton constant. We used the relation \cite{Gang:2019uay}
\begin{align}
\frac{L^2}{G_{(4)}} = \frac{2 N^3}{3\pi^2} \; \textrm{vol}(M_3) \;.
\end{align}
As we will show, this large $N$ free energy nicely matches with the  entropy function in \eqref{gravity entropy function}.

\subsubsection{Logarithmic subleading corrections}

The large $N$ logarithmic subleading correction --- proportional to $\log N$ --- is independent of the continuous deformation  parameter $\omega$. So we can compute it in the  $\omega \rightarrow 0$ limit, which is controlled by the overall factor $\bigl|\textrm{Hom}[\pi_1 (M_3) \rightarrow \mathbb{Z}_N] \bigr| / N$ and the perturbative expansion of $S^{\rm geom}_0$ and $S^{\rm geom}_1$. We have
\begin{align}
& \log \CI_{\rm sci} \bigl( q;\CT_N[M_3] \bigr) \Big|_{\omega \rightarrow 0,\, N\gg 1}  
\\
&= \frac{2}{i\omega} \, \textrm{Im}\bigl[ S^{\rm geom}_{0} (M_3;N) \bigr] + 2 \textrm{Re}  \bigl[ S^{\rm geom}_{1} (M_3;N) \bigr] + \log \frac{\bigl| \mathrm{Hom}[\pi_1 (M_3) \rightarrow \mathbb{Z}_N] \bigr| }{N} + \CO(\omega) \;.
\nn
\end{align}
From \eqref{large N of S0, S1} and (\ref{Selberg's expansion}) we see that there is no $\log N$ corrections from the first  two terms, $\textrm{Im}[S^{\rm geom}_0]$ and $\textrm{Re}[S^{\rm geom}_1]$.
The $\log N$ correction only comes from the last term, namely
\begin{equation}
\log \frac{\bigl| \mathrm{Hom}[\pi_1 (M_3) \rightarrow \mathbb{Z}_N] \bigr|}{N} =  \log \frac{\bigl| \mathrm{Hom}[H_1 (M_3,\mathbb{Z}) \rightarrow \mathbb{Z}_N] \bigr|}{N}
= \bigl( b_1(M_3)-1 \bigr) \log N\;.
\end{equation}
This matches the supergravity analysis in  \eqref{gravity log N} with $g=0$.  The comparison between the computation of logarithmic corrections in $SL(N,\mathbb{C})$ Chern-Simons theory and in supergravity is summarized in Table~\ref{computations of log N}.

\begin{table}[t]
	\begin{center}
		\begin{tabular}{|c|c|c|}
			\hline 
			& $SL(N,\mathbb{C})$ CS theory on $M_3$ & \rule[-.7em]{0pt}{2em} Supergravity on AdS$_4\times M_3 \times \widetilde{S}^4$ \\
			\hline\hline
			& \rule{0pt}{1.2em} \textrm{Quotient by} &  Two-form zero-modes on AdS$_4$ from
			\\
			$-\log N$ &residual gauge symmetry: & 2-form ghost field 
			\\ 
			&{$ \frac{1}{|\textrm{Stab}(\mathcal{A}^{\rm geom})|}  = \frac{1}{N} $}& in 11d supergravity
			\\[.5em]
			\hline 
			& \rule{0pt}{1.2em} Degeneracy of $\{\mathcal{A}^{\eta \otimes \alpha }\}$ from &  Two-form zero-modes on AdS$_4$ \\
			$b_1(M_3) \log N$  & tensoring with $\mathbb{Z}_N$ flat connections: & from 3-form potential $C_{(3)}$
			\\
			&$\bigl| \mathrm{Hom}[\pi_1 (M_3 )\rightarrow \mathbb{Z}^N] \bigr| \sim N^{b_1(M_3)}$ &   in 11d supergravity
			\\[.3em]
			\hline
		\end{tabular}
\caption {Comparison between the origin of logarithmic corrections to the superconformal index of $\mathcal{T}_N[M_3]$, from the 3d-3d dual $SL(N,\mathbb{C})$ Chern-Simons theory and from the holographic supergravity.}
		\label{computations of log N}
	\end{center}
\end{table}

\subsection[Twisted indices for $g>1$]{Twisted indices for \matht{g>1}}

A remarkable aspect of the 3d-3d relation for twisted indices \eqref{general 3d-3d for twisted index} is that only the one-loop  invariant $S_1^\alpha$ appears in the relation. Since we know the full perturbative $1/N$ expansion of the one-loop part, we can compute the full perturbative $1/N$ corrections to the twisted index of $\mathcal{T}_{N}[M_3]$ theory.  Using this idea, the full perturbative $1/N$ corrections were studied in \cite{Gang:2019uay}. In that work, the 3d-3d relation \eqref{3d-3d for twisted index} was limited to 3-manifolds with vanishing $H_1 (M_3, \mathbb{Z}_N)$.
With the more general 3d-3d relation for twisted indeces \eqref{general 3d-3d for twisted index}, we can repeat the large $N$ analysis for general closed hyperbolic 3-manifolds $M_3$. Interestingly, the logarithmic subleading correction $\log N$ can detect the first Betti number $b_1(M_3)$ of the 3-manifold. See \cite{Hosseini:2018qsx} for previous studies on full perturbative $1/N$ corrections to the twisted index of other classes of 3d theories.

\subsubsection[All orders  in $1/N$]{All orders  in \matht{1/N}}

Combining the 3d-3d relation  \eqref{general 3d-3d for twisted index} with the large $N$ behavior \eqref{large N of S0, S1}, we obtain for $g> 1$:
\begin{align}
& \mathcal{I}_{\Sigma_g} \bigl( \mathcal{T}_N[M_3] \bigr) 
\nn \\
&\xrightarrow{\; N\rightarrow \infty \;} \quad
\frac{N^{g-1}}{\bigl| \mathrm{Hom}[\pi_1(M_3) \rightarrow \mathbb{Z}_N]\bigr|^{g-1}} \; \biggl( e^{-2(g-1)\, S_1^\mathrm{geom}(M_3;N)} + \, c.c. \, + \, \ldots \biggr)
\\
&=\exp \biggl\{   (g-1)\frac{\textrm{vol}(M_3)}{6\pi } (2N^3-N-1)  - (g-1) \, \mathrm{Re} \sum_\gamma \sum_{s=1}^{\infty}\frac{1}s \left(\frac{e^{-s \ell_{\mathbb{C}}}}{1-e^{-s  \ell_{\mathbb{C}}}}\right)^2 
\nn \\
&\qquad \qquad -(g-1) \log \frac{\bigl| \mathrm{Hom}[\pi_1 (M_3) \rightarrow \mathbb{Z}_N] \bigr|}{N} +  \log \Bigl( 2 \cos\bigl( (g-1) \, \theta_{M_3,N} \bigr)\Bigr) \biggr\} + \ldots
\nn
\end{align}
Here the dots in the second line represent the contribution from all irreducible flat connections other than $\mathrm{geom}$ and $\overline{\mathrm{geom}}$. We assume that the contribution of non-geometric flat connections to $S_1$ is exponentially smaller than that of the geometric connections.%
\footnote{Such an assumption is justified, \textit{a posteriori}, because the geometric connections correctly reproduce the black hole entropy. We expect the single-center black hole to dominate the total entropy. Similar assumptions that one particular contribution dominates are made in other cases, including \cite{Benini:2015eyy}.}
The dots in the last line thus stand for exponentially smaller terms. The angle $\theta_{M_3,N} := 2 \, \mathrm{Im}[S_1^\mathrm{geom}]$ represents the phase factor from summing over the Bethe vacua $\alpha = {\rm geom}$ and  $\alpha = {\overline{\rm geom}}$.
For $M_3$ with vanishing $H_1(M_3, \mathbb{Z}_N)$, the large $N$ analysis was already done in \cite{Gang:2019uay} and it was checked that the leading $\CO(N^3)$ term nicely matches the entropy of magnetically-charged universal black holes in the holographic dual AdS$_4$. 
The logarithmic correction is 
\begin{align}
- (g-1) \log \frac{\bigl| \mathrm{Hom}[\pi_1 (M_3) \rightarrow \mathbb{Z}_N] \bigr|}{N} \quad \xrightarrow{\, N\rightarrow \infty \, } \quad (g-1)\big{(}1-b_1 (M_3)\big{)} \log N\;.
\end{align}
This also matches the supergravity analysis in \eqref{gravity log N}.

\subsection[Refined  index: no $N^3$ behavior]{Refined  index: no \matht{N^3} behavior}

From \eqref{3d-3d for refined index}, \eqref{large N of S0, S1} and \eqref{large N of Sn n>=2} we obtain
\begin{align}
&\CI_{\rm top}\bigl( q;\CT_{N}[M_3] \bigr)  =     \frac{\bigl| \mathrm{Hom}[\pi_1 (M_3) \rightarrow \mathbb{Z}_N] \bigr|}{N}   \sum_{[\mathcal{A}^\alpha] \in \frac{\chi_{\rm irred} (M_3;N)}{\mathrm{Hom}[\pi_1 (M_3) \rightarrow \mathbb{Z}_N]} }B^\alpha (q) \; B^{\alpha}(q^{-1}) 
\nn \\
& =  \frac{\bigl| \mathrm{Hom}[\pi_1 (M_3) \rightarrow \mathbb{Z}_N] \bigr|}{N} \; B^{\rm geom} (q) \, B^{\rm geom}(q^{-1})  + \, c.c. \, + \; \bigl( \textrm{other $\alpha$'s} \bigr)
\nn \\
&=   \frac{\bigl| \mathrm{Hom}[\pi_1 (M_3) \rightarrow \mathbb{Z}_N] \bigr|}{N}  \;  \exp \left(2 \sum_{n=0}^\infty (-\omega)^{2n} S_{2n+1}^{\rm geom} \right)    + \; \ldots
\\
&=   \frac{\bigl| \mathrm{Hom}[\pi_1 (M_3) \rightarrow \mathbb{Z}_N] \bigr|}{N}  \;   \exp \left( - \frac{2 N^3 \textrm{vol}(M_3)}{3\pi} + \, \textrm{subleading} \right) + \; \ldots \;.
\nn
\end{align}
As before, the dots stand for contributions from other $\alpha$'s than the geometric ones. We see that for the refined index, unlike for the superconformal index, the contributions from $\mathcal{A}^{\rm geom}$ and $\mathcal{A}^{\overline{\rm geom}}$ are exponentially small in the large $N$ limit. This is compatible with the curious observation we made in section~\ref{subsec : integrality for refined index} that the refined twisted index seems to actually vanish at finite $N$. 
This is also compatible with the non-existence of magnetically-charged  black holes with AdS$_2\times S^2$ horizon in the universal sector, as will be discussed in subsection~\ref{sec : no uni mag spinning BH}.

%%%%%%%%%%%%%%%%%%%%%%%%%%%%%%%%%%%%%%%%%%%%%%%%%%%%%%%

\section{M5-branes wrapped on hyperbolic \matht{M_3} and \matht{{\cal N}{=}2} gauged supergravity}
\label{Sec:Sugra}

To motivate  the holographic description of the field theories discussed in the previous sections, we start in eleven dimensions where the M5-brane naturally resides. Since the AdS/CFT dictionary identifies the conformal algebra in field theory with the isometries of spacetime on the gravity side, we expect a gravitational solution given by a backreacted AdS solution in which M5-branes are partly wrapped on $M_3$.

The study of various solutions representing the supergravity description of M5-branes wrapping hyperbolic 3-manifolds has a distinguished history starting with \cite{Gauntlett:2000ng}. Although much of the emphasis has been on the construction of explicit solutions, starting with AdS$_4$ vacuum solutions and expanding into more general black hole solutions, there are fruitful efforts in the direction of consistent truncations \cite{Gauntlett:2007ma, Donos:2010ax, Gang:2014ema}.
Our strategy to describe the consistent truncation is to present the AdS$_4$ vacuum solution in the 11d context  and then to indicate how the setup generalizes to yield a consistent truncation of 11d supergravity to 4d minimal ${\cal N}=2$ gauged supergravity with just the graviphoton field.

The central intuition for understanding the 11d origin of M5-branes wrapping hyperbolic 3-manifolds is provided by the Pernici-Sezgin solution to 7d gauged supergravity originally constructed in  \cite{Pernici:1984nw}: roughly it discusses solutions of the type AdS$_4\times M_3$. The original 11d point of view was discussed in \cite{Gauntlett:2006ux},%
\footnote{See also \cite{Gabella:2012rc} for a more geometric $SU(2)$-structure-based approach, and \cite{Bah:2014dsa} for a concise exposition.}
while the 7d gauged supergravity point of view can be recovered as a consistent reduction.
In 11d the solution has the form AdS$_4\times Y_7$, where $Y_7$ is an $S^4$ fibration over the  hyperbolic 3-manifold $M_3$ \cite{Gauntlett:2006ux}:
\begin{align}
ds^2_{11} &= \lambda^{-1} \Bigl( ds^2(\mathrm{AdS}_4) + ds^2_7 \Bigr) \\
&= \lambda^{-1}ds^2(\mathrm{AdS}_4)+ds_4^2 \bigl( {\cal M}_{SU(2)} \bigr) + \hat{w}\otimes \hat{w}+\frac{\lambda^2}{16}\left(\frac{d\rho^2}{1-\lambda^3\rho^2}+\rho^2d\psi^2\right) \;,
\nn
\end{align}
where $\lambda$ is the warp factor, ${\cal M}_{SU(2)}$ is a 4d space with $SU(2)$ structure, $\hat{w}$ is a one-form, $\rho$ is a coordinate on an interval, and $\psi\in [0,2\pi]$ is a coordinate on the circle $S^1$.  The associated Killing vector $\partial/\partial \psi$ is dual to the $U(1)$ R-symmetry in field theory.

The particular form of the Pernici-Sezgin solution \cite{Pernici:1984nw} is (see \cite{Gauntlett:2006ux}):
\be
ds_4^2\bigl( {\cal M}_{SU(2)} \bigr) + \hat{w}\otimes \hat{w} = f^2(\rho) \, DY^a \otimes DY^a + g^2(\rho) \, e^a \otimes e^a \;,
\ee
where $Y^a$ ($a=1,2,3$) are constrained coordinates on $S^2$ satisfying $Y^a Y^a=1$, whilst $e^a$ is a vielbein on the 3-manifold $M_3$.%
\footnote{We follow our convention and normalize the metric on $M_3$ in such a way that the Ricci scalar is $R=-6$.}
The form of the covariant derivative is  
\be
DY^a =dY^a +\omega^a {}_b Y^b \;,
\ee
where $\omega^{ab}$ is the spin connection on $M_3$.  The supersymmetry conditions dictate
\be
\lambda^3=\frac{2}{8+\rho^2} \;,\qquad f=\frac{\sqrt{1-\lambda^3\rho^2}}{2\sqrt{\lambda}} \;,\qquad g=\frac{1}{2\sqrt{\lambda}}
\ee
and the coordinate $\rho$ covers the interval $\bigl[ 0, 2\sqrt2 \bigr]$.
The coordinates $Y^a, \rho, \psi$ would form a round $S^4$ in the undeformed case, but here they form instead a space we denote by $X_4$.

A concrete way to realize the hyperbolic manifold $M_3$ is to take a quotient of the hyperbolic plane $\mathbb{H}^3$ by a discrete subgroup $\Gamma\in PSL(2,\mathbb{C})$, namely $M_3 = \mathbb{H}^3/\Gamma$. The background is then
\begin{align}
\label{Sol:AdS4}
ds_{11}^2 &= \frac1{4\lambda} \bigg[4 \, ds^2_\textrm{AdS$_4$} + \frac{dx^2+dy^2+dz^2}{z^2} + \frac{8-\rho^2}{8+\rho^2} (DY^a)^2 +
\frac{1}{2}\left(\frac{d\rho^2}{8-\rho^2}+\frac{\rho^2}{8+\rho^2}d\psi^2\right)\bigg] \nonumber \\
G_4 &= \frac{1}{4} \, d\psi\wedge d\left(\lambda^{-1/2}\sqrt{1-\lambda^3 \rho^2} \, J_3\right) \;.
\end{align}
A slightly more general class of solutions involves the standard two-forms $(J_1, J_2, J_3)$ defining the $SU(2)$ structure.
The ansatz for this class of solutions is \cite{Gauntlett:2006ux}
\begin{align}
\hat{w} &= g Y^a e^a \;, \nonumber \\
J_1 &= f\, g \, DY^a\wedge e^a \;, \nonumber \\
J_2 &= f\, g\, \epsilon^{abc} \, Y^a DY^b\wedge e^c \;, \nonumber \\
J_3 &= \tfrac{1}{2} \, \epsilon^{abc} \, Y^a\left(f^2 DY^b \wedge DY^c -g^2 e^b \wedge e^c\right) \;.
\end{align}

To understand the quantization condition of the flux and the identification with the number of M5-branes, we need to consider the  four-form field $G_4$ of 11d supergravity when restricted to the internal space $X_4$. In fact, the $G_4$ flux in (\ref{Sol:AdS4}) is entirely aligned along $X_4$.
The free energy can be easily computed \cite{Gabella:2012rc} to yield%
\footnote{We correct a typo in eqn. (2.9) of \cite{Bah:2014dsa}.}
\be
\mathcal{F} = \frac{4\pi^3 \int_{Y_7} \lambda^{-9/2} \, d\mathrm{vol}_7 }{(2\pi l_P)^9} =
\frac{\pi^5 \, \mathrm{vol}(M_3)}{3\,(2\pi l_P)^9} \;.
\ee
The 4-form flux quantization leads to a relation between Planck's constant $l_P$ and $N$:
\be
N=\frac{1}{(2\pi l_P)^3}\int_{X_4}G_4=\frac{1}{8\pi l_P^3} \;.
\ee
Substituting, we can write the free energy in terms of field theory parameters: 
\be
{\cal F}=\frac{N^3}{3\pi} \mathrm{vol}(M_3) \;.
\ee
This result was generalized to account for the free energy corresponding to a squashed 3-sphere $S_b^3$ in \cite{Martelli:2011fu, Gang:2014qla}:
\be
{\cal F}_b=\frac{1}{4}\left(b+\frac{1}{b}\right)^2 {\cal F} \;.
\ee

The steps to obtain the 11d embedding of 4d minimal ${\cal N}=2$ gauged supergravity are intuitively clear \cite{Gauntlett:2007ma}.%
\footnote{See also \cite{Larios:2019kbw, Larios:2019lxq, Varela:2019vyd} for other consistent truncations to minimal $\cN=2$ gauged supergravity.}
Having identified the  Killing vector $\partial/\partial \psi$ as the dual to the $U(1)$ R-symmetry, we simply shift $d\psi \to  d\psi -A$ in the metric given in equation (\ref{Sol:AdS4}), where $A$ is the 4d one-form gauge field. The general ansatz for $G_4$ is somehow more involved, containing both the field strength $F=dA$ as well as its Hodge dual; it can be found in \cite{Gauntlett:2007ma}.
For our purpose the most important property is that this generalization does not affect the quantization condition above. 

The Einstein-graviphoton part of the action is simply
\begin{equation}
\label{Eq:4dGrav}
I = \frac{1}{16\pi G_{(4)} }\int d^4x \, \sqrt{-g} \, \left( R + \frac{6}{L^2} - \frac{L^2}4 F^2 \right) \;,
\end{equation}
where $G_{(4)}$ is the 4d Newton constant --- not to be confused with $G_4$.
Note that in the expression above we have explicitly restored the radius of the AdS vacuum solution, which we had set to one in previous expressions. 
The theory given by (\ref{Eq:4dGrav}) is precisely the universal sector discussed  recently in  \cite{Azzurli:2017kxo} in the context of microscopic counting of AdS$_4$ black hole entropy. Here the crucial property is the embedding into M-theory and, more precisely, the scaling of Newton's constant with the number of branes $N$. 
In the action above, $F$ is the field strength for the $U(1)$ gauge field in AdS$_4$ that couples to the $U(1)$ R-symmetry of the boundary CFT. The 4d Newton constant $G_{(4)}$ after the consistent truncation is related to $N$ in the following way \cite{Gang:2014ema}:
\begin{equation}
\frac{G_{(4)}}{L^2} =\frac{3 \pi^2}{2 N^3 \, \mathrm{vol}(M_3)} \;. 
\end{equation}
This is compatible with the general result $\mathcal{F} = \pi L^2/2 G_{(4)}$ \cite{Gabella:2012rc}.

%%%%%%%%%%%%%%%%%%%%%%%%%%%%%%%%%%%%%%%%%%%%%%%%%%%%

\subsection[Rotating electrically-charged  AdS$_4$ black hole]{Rotating electrically-charged  AdS\matht{_4} black hole}

The original placement of the rotating solutions in the context of gauged ${\cal N}=2$ supergravity was presented in  \cite{Kostelecky:1995ei,Caldarelli:1998hg}. In this manuscript we follow the notation of \cite{Cvetic:2005zi} in which the background takes the following form:
\begin{align}
\label{Eq:Solution}
ds^2 &= -\frac{\Delta_r}{W} \biggl( dt -\frac{a}{\Xi} \sin^2\theta \, d\phi \biggr)^2+W\left(\frac{dr^2}{\Delta_r}+\frac{d\theta^2}{\Delta_\theta}\right)
+\frac{\Delta_\theta \sin^2\theta}{W}\left(a \, dt -\frac{\rho^2+a^2}{\Xi} \, d\phi \right)^2 \nonumber \\
\rho &= r+2m \sinh^2\delta \nonumber\\
\Delta_r &= r^2 +a^2 -2mr +g^2 \rho^2 (\rho^2 +a^2) \qquad\qquad \Delta_\theta = 1-a^2 g^2 \cos^2\theta \nonumber \\
W &= \rho^2 +a^2 \cos^2\theta \hspace{12.2em} \Xi =1-a^2 g^2 \nonumber \\
A &= \frac{2m \sinh2\delta}{W} \, \rho \, \left(dt -\frac{a}{\Xi}\sin^2\theta \, d\phi\right) \;.
\end{align}
The parameter $g$ simply characterizes the strength of the coupling in gauged supergravity and is inversely proportional to the radius of AdS, $g= L^{-1}$. 
The physics of this solution is parametrized  by $(m, a, \delta)$ which roughly characterize the energy, the angular momentum and the non-extremality or charge of the solution. More precisely:%
\footnote{In \cite{Cvetic:2005zi} Newton's constant $G_{(4)}$ was set to $1$.}
\begin{equation}
E = \frac m{G_{(4)} \Xi^2} \cosh 2\delta \;,\qquad J = \frac{m\,a}{G_{(4)} \Xi^2} \cosh 2\delta \;,\qquad Q = \frac{m}{4 G_{(4)} \Xi} \sinh 2\delta \;.
\end{equation}

This solution has been, of course, generalized to include its embedding in 4d ${\cal N}=8$ gauged supergravity with gauge group $U(1)^4$.  This fact makes it a prime object in the study of the AdS/CFT correspondence due to its relation with ABJM theory \cite{Aharony:2008ug}. When the four charges with respect to $U(1)^4$ are set equal pairwise, one obtains a solution which is naturally embedded into 4d ${\cal N}=4$ gauged supergravity with gauge group $U(1)\times U(1)$. This is the solution that has been widely discussed in \cite{Cvetic:2005zi}, focusing on the supersymmetric limit of the non-extremal solution originally presented in \cite{Chong:2004na}. Technically,  in this paper we are concerned with the limit of the solution discussed in \cite{Cvetic:2005zi} when $\delta_1=\delta_2$ (in their notation) which we simply denote by $\delta$. Versions of these solutions have recently been reviewed in  \cite{Choi:2018fdc, Cassani:2019mms} with emphasis on the entropy function. 

Having a potential microscopic description of these solutions, we are naturally interested in the Bekenstein-Hawking entropy $S_\textrm{BH}$ in the BPS limit where one verifies that \cite{Choi:2018fdc, Cassani:2019mms}
\be
S_\textrm{BH} = \frac{\pi L}{4 G_{(4)}} \; \frac{J_\textrm{BPS}}{Q_\textrm{BPS}} \;.
\ee
Due to supersymmetry and extremality, $J_\textrm{BPS}$ and $Q_\textrm{BPS}$ are non-linearly related.
The entropy can then more explicitly be written as:
\be
S_\textrm{BH} = \frac{\pi L^2}{2G_{(4)}} \left( \sqrt{1 + 64 \frac{G^2_{(4)}}{L^2} Q_\textrm{BPS}^2} \;\; - 1 \right) \;.
\ee
The gravitational entropy function $\mathcal{S}$ is the ``grand canonical partition function'' that reproduces the Bekenstein-Hawking entropy upon Legendre transform:
\be
S_\textrm{BH}\big(Q_\textrm{BPS} \bigr) = \mathcal{S}(\omega) + \omega \, J_\mathrm{BPS} + (\omega + 2\pi i) \, 2L Q_\mathrm{BPS}
\ee
where $\omega(Q_\textrm{BPS})$ is fixed by requiring that $\partial(\mathrm{RHS})/\partial\omega = 0$. The entropy function $\CS$ also reproduces the non-linear constraint between $J_\textrm{BPS}$ and $Q_\textrm{BPS}$, as the condition that $S_\textrm{BH}$ be real positive \cite{Benini:2016rke, Cabo-Bizet:2018ehj}. The entropy function can also be computed from the on-shell action of the solution. In any case, it takes the following form \cite{Choi:2018fdc, Cassani:2019mms}:
\be
\label{gravity entropy function}
\mathcal{S} = i \, \frac{L^2}{8G_{(4)}} \, \frac{(\omega + 2\pi i)^2}{\omega} = \frac{i \, N^3}{12\pi^2} \; \mathrm{vol}(M_3) \; \frac{(\omega+2\pi i )^2}{\omega} \;,
\ee
where we used the explicit form of the four-dimensional Newton constant $G_{(4)}$. This expression perfectly matches the field theory computation (\ref{final result SCI}). This justifies the use of the field theory description as a microscopic explanation for the entropy of the rotating electrically-charged asymptotically AdS$_4$ black holes described above.

%%%%%%%%%%%%%%%%%%%%%%%%%%%%%%%%%%%%%%%%%%%%%%%%%%

\subsection[Rotating magnetically-charged AdS$_4$ black holes]{Rotating magnetically-charged AdS\matht{_4} black holes}
\label{sec : no uni mag spinning BH}

There have been many studies regarding rotating magnetically-charged black holes that are the natural candidates for the refined topologically twisted index. In particular, there is  one recent publication that discusses these issues in detail   and contains a comprehensive list of references \cite{Hristov:2018spe}. A common theme, since the  very beginning of the explorations of rotating magnetically-charged black holes, is that within the confines of the universal sector described by the action (\ref{Eq:4dGrav}), spherically symmetric black holes do not exist \cite{Caldarelli:1998hg}. Indeed, the need to couple to matter was recognized very early on \cite{Klemm:2011xw}.  

The absence of such supergravity solutions is compatible with the result on the field theory side that the refined topologically twisted index vanishes. It would be interesting  to explored the refined topologically twisted index for the hyperbolic two-plane, in which case there are supergravity solutions that one might hope to match.

%%%%%%%%%%%%%%%%%%%%%%%%%%%%%%%%%%%%%%%%%%
\section{Logarithmic corrections from one-loop supergravity}
\label{Sec:OneLoop}

Logarithmic corrections to the extremal black hole entropy can be computed purely in terms of the low energy data, that is, from the spectrum of massless fields in a given gravitational background. These IR corrections are expected to be reproduced by any candidate to a UV complete description of the gravitational theory. Such an IR window into UV physics has been exploited by Ashoke Sen and collaborators in the context of asymptotically flat black holes whose microscopic description lies in string theory \cite{Sen:2011ba, Sen:2012cj}. For asymptotically AdS black holes  the microscopic UV complete theory is provided by the dual boundary field theory. It is, therefore, natural to study if the terms logarithmic in $N$ coming from the field theory partition functions are reproduced by one-loop supergravity around the black hole backgrounds. There have been some developments in matching the gravity computation to the coefficient of the $\log N$ term on the field theory side \cite{Liu:2017vll, Jeon:2017aif, Liu:2017vbl, Hristov:2018lod, Liu:2018bac, PandoZayas:2019hdb, Hristov:2019xku, Gang:2019uay}. In this section we compute the one-loop logarithmic corrections from the gravity side and confront them with the field-theoretic (UV) results. One of the advantages of working in the context of the 3d-3d correspondence is that the logarithmic term is derived analytically, without the recourse to numerical methods as was instead the case in \cite{Liu:2017vll, Liu:2018bac, PandoZayas:2019hdb}.

Let us start by recalling a number of important facts regarding one-loop effective actions on supergravity backgrounds. Our setup is 11d supergravity, in which we have an explicit embedding of the solutions describing rotating black holes. We highlight  that in odd dimensional spaces only zero-modes and boundary terms can contribute to logarithmic terms. We make the assumption that the whole contribution to the action comes from the asymptotic AdS$_4$ region, as it was the case in \cite{Bhattacharyya:2012ye} for the pure AdS$_4$ solution and in \cite{Liu:2017vbl} for the magnetically-charged asymptotically AdS$_4$ black hole solution. More directly related to our description of M5-branes wrapping hyperbolic 3-manifolds is the recent discussion in \cite{Gang:2019uay}. 

On very general grounds of diffeomorphic invariance, it can be argued that in  odd-dimensional spacetimes the top Seeley-De Witt coefficient $a_{d/2}$ vanishes \cite{Vassilevich:2003xt}. Therefore, the only contribution to the heat kernel comes from zero-modes. Applied to our case, the one-loop contribution in 11d supergravity comes from the analysis of zero-modes. Similar properties have, in fact, been already exploited in the context of the logarithmic corrections to BMPV black holes in  \cite{Sen:2012cj}, where logarithmic contributions can be understood through an effective 5d theory. Similarly, the authors of \cite{Bhattacharyya:2012ye} successfully matched the logarithmic term in the large $N$ expansion of the ABJM free energy on $S^3$ with a gravity computation  performed in 11d supergravity, essentially reducing to the contribution of a two-form zero-mode.  Along these lines other matches for magnetically-charged asymptotically AdS$_4$ black holes, dual to the topologically twisted index, were performed in \cite{Liu:2017vbl, Gang:2019uay}.

We will not reproduce all the details of the computation here: the interested reader is referred to \cite{Bhattacharyya:2012ye, Liu:2017vbl, Gang:2019uay} for details. Instead, we will briefly sketch the derivation of the one-loop effective action. Given that the only zero-mode on AdS$_4$ is a 2-form, and assuming that the solution is roughly of the form of a warped product AdS$_4 \times M_3\times \widetilde{S}^4$, we need to decompose the kinetic operator along those three subspaces. In order for the 2-form zero-mode on AdS$_4$ to survive, we need to have the corresponding part of the kinetic Laplace-like operator also vanishing. The number of zero-modes then depends on the topology of the full space. 

When integrating over zero-modes, there is a  factor of $L^{\pm \beta_A}$ for each zero-mode, where $\beta_A$ is a coefficient to be determined. The total contribution to the partition function from the zero-modes is then
\bea
L^{\pm \beta_A \: n_A^0} \;,
\eea
where $n_A^0$ is the total number of zero-modes of the kinetic operator $A$, and the sign depends on whether the operator is fermionic or bosonic. Typically, zero-modes are associated with certain asymptotic symmetries --- for example, with gauge transformations that do not vanish at infinity. The key idea in determining $\beta_A$ is to find the right integration variables and to count the number of powers of $L$ that such integration measure contributes, when one starts from fields that would naturally be present in the action.   

Therefore, in our case, the computation of the one-loop effective action reduces to the computation of the zero-mode scaling $\beta_A$ and the total number of zero-modes $n_A^0$. The most important ingredient in formulating the answer is the number of 2-form zero modes.  A simple way to determine their number is by computing the Euler characteristic of the black hole. In \cite{Liu:2017vbl, Gang:2019uay}  it was established that $n_2^0=2(1-g)$ for a black hole whose horizon is a genus $g$ Riemann surface. It is worth pausing over this result. Note that this number is computed using the non-extremal branch of the solution. Moreover, it is independent of the black hole charges. Therefore, be it for the magnetically-charged or the electrically-charged black holes, we obtain the same result. 

The last step before writing down the answer is to recall that to quantize eleven-dimensional supergravity we need to discuss the quantization of the 3-form field, $C_3$. The action for quantizing a $p$-form $A_p$ requires a set of $(p-j)$-form ghost fields, with $j=1, \ldots, p$, and the ghost is Grassmann even (odd) if $j$ is even (odd) \cite{Siegel:1980jj,Copeland:1984qk}. For the $(p-j)$-form, the Laplacian operator $(\Delta_{p-j})^{j+1}$ in the computation of the heat kernel leads to an extra scaling power $-(j+1)$, but otherwise the integration over the zero-modes is unchanged. The result, as in eqn.~(3.4) of \cite{Bhattacharyya:2012ye}, is
\begin{equation}
\Delta F_{\text{Ghost}}=\sum_{j=0}^p (-1)^j \, \bigl( \beta_{A_{p-j}} - j -1 \bigr) \, n_{A_{p-j}}^0 \log L \;.
\end{equation}
The scaling exponent for $p$-forms is easily computed \cite{Bhattacharyya:2012ye}, yielding $\beta_p = (d-2p)/2$ in terms of the total dimension $d$ of spacetime.

The zero-mode contribution in our solution can only come from a ghost 2-form $A_{p-j=2}$ which corresponds to $p=3$ and $ j=1$ and it leads to the following one-loop contribution 
\bea
\label{zero-mode contribution}
(-1)^j \bigl(\beta_{2} - j-1 \bigr) \, n^0_{2} \, \log L = \bigl( 2-\beta_{2} \bigr) \, n^0_2 \, \log L \;.
\eea
The full contribution to the logarithmic terms of the one-loop effective action is thus given only by the  2-form zero-modes and we have:
\be
\log Z_\mathrm{1-loop} = (2-\beta_2) \, n_2^0 \, \log L= \bigl(2-7/2 \bigr) \, 2(1-g) \, \log L = (g-1) \log N \;,
\ee
where, according to the structure of the M5-brane solution, we have $L^3\sim N$. When restricting to spherically symmetric horizons $(g=0)$  we find perfect agreement with the field theory result. 

In \cite{Gang:2019uay} the case when $M_3$ admits harmonic one-forms was further discussed. In this case there are extra contributions coming from the supergravity 3-form potential, because one can construct a zero-mode of $C_3$ by combining a 2-form zero-mode on AdS$_4$ with a harmonic 1-form on $M_3$. The contribution is as in (\ref{zero-mode contribution}), but with $p=3$, $j=0$, $\beta_3 = 5/2$ and the total number of zero-modes $n_3^0 = 2b_1(1-g)$ where $b_1$ is the first Betti number of $M_3$. Thus, the total contribution to the one-loop partition function takes the form:%
\footnote{This expression corrects a sign error in \cite{Gang:2019uay}.}
\be
\label{gravity log N}
\boxed{\quad\rule[-.7em]{0pt}{2em}
\log Z_\textrm{1-loop} = (g-1) (1-b_1) \log N \;.
\quad}
\ee
This results perfectly agrees with the field theory computation of the coefficient of the term logarithmic in $N$. It thus provides strong evidence that the superconformal index captures the black hole entropy beyond leading order.

%%%%%%%%%%%%%%%%%%%%%%%%%%%%%%%%%%%%%%%%%%

\section{Conclusions}
\label{Sec:Conclusions}

In this manuscript we have presented an explicit recipe for, and the computation of, the superconformal index of  ${\cal N}=2$ supersymmetric field theories denoted by  $ \mathcal{T}_N [M_3] $. Our recipe identifies the superconformal index  with certain pertubative invariants of  $SL(N, \mathbb{C})$ Chern-Simons theory  via the 3d-3d correspondence.  One advantage of this setup is that it provides analytical results for the leading term and even for the term logarithmic in $N$. Recent approaches to the microscopic counting of states of rotating electrically-charged asymptotically AdS$_4$ black holes have relied on reducing the field-theoretic object to a matrix model, where the leading order can be easily found but the  subleading in $N$ corrections seem like a daunting task   \cite{Choi:2019zpz, Nian:2019pxj}. We showed that the leading contribution to the superconformal index perfectly matches the entropy of rotating electrically-charged black holes in the universal sector of ${\cal N}=2$ gauge supergravity. More importantly, we demonstrated that the terms logarithmic in $N$ also match on both sides of the AdS/CFT correspondence. 

We have also studied the refined topologically twisted index and found agreement between its vanishing on the field theory side and the absence of dual rotating magnetically-charged black holes in the universal sector of ${\cal N}=2$ gauged supergravity. Our approach provides strong evidence in favour of the refined index vanishing also for finite $N$. This observation has powerful mathematical implications, and it would be quite interesting to pursue this tantalizing relation between number-theoretic objects and black hole physics. 

Our subleading result is a powerful tool that could be applied to other situations where various observables are vying for the genuine description of black hole microstates.  The situation with various, {\it a priori}, different observables yielding the same leading order expression for the black hole entropy  in  AdS$_4$ is quite similar to the situation  in AdS$_5$.  In the context of rotating electrically-charged asymptotically AdS$_5$ black holes there  are various observables that, at leading order, yield the same entropy function and, therefore, compete for the microscopic description of the entropy.   These  observables include: a localization computation on complex backgrounds \cite{Cabo-Bizet:2018ehj}, the  superconformal index computation via Bethe Ansatz \cite{Benini:2018mlo, Benini:2018ywd} and a free-field theory approach \cite{Choi:2018hmj}.  It is quite plausible that these observables are equivalent; it is also possible that a subleading in $N$ study has the chance to break this degeneracy. It would be quite appropriate to study this problem in more detail.

\subsection{Future directions}

Let us mention a few open questions that would be nice to address.

\paragraph{Curious observation \matht{\mathcal{I}_\mathrm{top}\bigl( q; \mathcal{T}_N[M_3] \bigr) =0}.}
We checked that $\mathcal{I}_\mathrm{top}\bigl( q; \mathcal{T}_N[M_3] \big) =0$ to some high power in $\hbar$, for some examples of closed hyperbolic 3-manifold $M_3$. Is this just a coincidence? 
It would be  interesting to check if the vanishing holds for arbitrary closed hyperbolic 3-manifolds. A rigorous mathematical proof of the vanishing for infinitely many closed hyperbolic 3-manifolds at $q=1$ will be reported in \cite{Gang:2019dbv}. 
 If the answer is positive, we need to understand such a non-trivial property either from  $SL(N,\mathbb{C})$ Chern-Simons theory or from the physics of wrapped M5-branes. In this paper we have offered a physical argument from the viewpoint of wrapped M5-branes.

\paragraph{Going beyond the universal sector in the 3d-3d correspondence.}
In this paper we only considered the 3d-3d correspondence for closed 3-manifolds $M_3$ without any codimension-two defect. The resulting 3d $\mathcal{N}=2$ theory (at sufficiently large $N$) only has $SO(2)$ R-symmetry without any flavor symmetry. Holographic duals for these theories  are well-established \cite{Pernici:1984nw, Gauntlett:2000ng}, and allowed us to perform one-loop computations on the gravity side. The system, however, could be generalized by introducing codimension-two defects in the 3d-3d setup \eqref{3d-3d set-up}. 
Indeed, the  6d $A_{N-1}$ $\mathcal{N}=(2,0)$ theory has regular supersymmetric codimension-two defects. Putting those defects along $(\textrm{3d space-time})\times (\textrm{a knot  $\CK$ inside $M_3$})$, we can engineer more general classes of 3d theories labelled by $(M_3, \CK)$ and  the types of the regular defects. We have concrete field theoretic constructions of the 3d theory  only when the codimension-two defect is of ``maximal type'' \cite{Dimofte:2011ju, Dimofte:2013iv}. In order to extend our work and to understand more general classes of supersymmetric black holes in AdS$_4$, we need a better understanding on the codimension-two defects in  the context of the 3d-3d correspondence, both from the gauge theory side and  the supergravity side. It would be quite interested to pursue this direction, elaborating on the progress reported in \cite{Bah:2014dsa, Gang:2015wya}.

\paragraph{Other geometries.}
From the point of view of the 3d $\cN=2$ theories $\CT_N[M_3]$, there are other types of supersymmetric partition functions one could compute, for instance the lens space index \cite{Benini:2011nc} or the partition function on Seifert manifolds \cite{Closset:2018ghr}. In all those cases, suitable 3d-3d relations could be used to relate supersymmetric partition functions to (possibly Euclidean) supergravity backgrounds.

%%%%%%%%%%%%%%%%%%%%%%%%%%%%%%%%%%%%%%%%%%%%%%%%%%%%%%

\section{Acknowledgments}

We are thankful to Sergio Cecotti, Francesca Ferrari, Sergei Gukov, Saebyeok Jeong, Heeyeon Kim, Jun Nian, Pavel Putrov, Chiara Toldo and  Kazuya Yonekura for helpful discussions.
This work was partially done while one of the authors (DG) was visiting APCTP, Pohang for the workshop ``Strings, Branes and Gauge Theories'', 17--27 July 2019. We thank APCTP for hospitality.
FB is supported in part by the MIUR-SIR grant RBSI1471GJ  ``Quantum Field Theories at Strong Coupling: Exact Computations and Applications''.
The research of DG  is supported in part by the National Research Foundation of Korea under grant 2019R1A2C2004880. DG also acknowledges support by the appointment to the JRG program at the APCTP through the Science and Technology Promotion Fund and Lottery Fund of the Korean Government, as well as support by the Korean Local Governments, Gyeongsangbuk-do Province, and Pohang City.
The work of LAPZ is supported in part by the U.S. Department of Energy under grant DE-SC0007859.

%%%%%%%%%%%%%%%%%%%%%%%%%%%%%%%%%%%%%%%%%%%%%%%%%%%%%%
\appendix

\section{Perturbative invariants \matht{S_n^{\alpha}[M_3; N=2]} for some \matht{M_3}'s}
\label{App: perturbative invariants}
\label{App:Examples}

\paragraph{Dehn-surgery representation.}
Every orientable closed 3-manifold can be obtained by Dehn surgery along a link $L$ inside the three-sphere $S^3$. The 3-manifold $M_3$ obtained by a Dehn surgery along a link $L$ with slopes $\{P_i/Q_i\}_{i=1}^{|L|}$ is defined as
\begin{align}
\label{Dehn surgery representation}
& M_3 = \bigl( S^3\backslash L \bigr)_{P_1 /Q_1,\, \ldots \,, P_{|L|}/Q_{|L|}}\;
\\
&\quad\; :=\left[ \bigl(S^3\backslash L \bigr) \, \bigcup \, \bigl( |L| \; \textrm{solid-tori} \bigr)  \right]/\sim \quad \textrm{where the gluing is determined  by the slopes}
\nn \\
& P_i \mu_i  + Q_i \lambda_i  \, \in\, H_1 \bigl( \partial(S^3\backslash L),\mathbb{Z}\bigr) \sim (\textrm{shrinkable boundary cycle of $i$-th solid torus})\;.
\nn
\end{align}
Here $|L|$ is the number of components of the link $L$ and $S^3\backslash L$ denotes the link complement obtained from removing a tubular neighborhood of the link $L$ from $S^3$. The boundary of the link complement consists of $|L|$ tori. For each boundary torus, there is a canonical choice of basis 1-cycles, called merdian $\mu$ (bounding a disk in the tubular neighborhood of the link) and longitude $\lambda$ (homologous to the link component). In particular, $\mu_i$ and $\lambda_i$ are the merdian and the longitude of the $i$-th boundary torus, respectively. Then $P_i$ and $Q_i$ are coprime integers and $P_i/Q_i \in \mathbb{Q} \cup \{\infty \}$ is called the $i$-th slope of Dehn surgery. The procedure of gluing solid tori back to the link complement is called {\it Dehn filling}.
After Dehn filling, the resulting manifold $M_3$ is a closed 3-manifold without any boundary.

\paragraph{Perturbative invariants from state-integral model.}
Using the topological nature of  Chern-Simons theory, its path-integral can be reduced to  a finite dimensional integral  which is called a state-integral model. 
See \cite{Dimofte:2009yn, Dimofte:2012qj, Gang:2017cwq, HIKAMI_2001, Hikami_2007, Dimofte:2011gm, Bae:2016jpi} for state-integral models of $SL(N,\mathbb{C})$ Chern-Simons theory from which the perturbative invariants  $\bigl\{S_n^{\alpha}[M_3;N] \bigr\}$  can be efficiently computed. 
The state-integral model turns out to be equal to the localization integral of $\mathcal{T}_{N}[M_3]$ on a squashed 3-sphere $S^3_b$ \cite{Hama:2011ea} with the identification $\hbar = 2\pi i b^2$.  Here we give concrete expressions for the state-integral models at $N=2$ corresponding to the three smallest closed hyperbolic 3-manifolds, namely Weeks manifold ${} = (S^3\backslash \mathbf{5}^2_1)_{P_1/Q_1 = -5 ,\, P_2/Q_2 =-5/2 }$, $(S^3\backslash \mathbf{4}_1)_{P/Q=5}$ and $(S^3\backslash \mathbf{5}^2_1)_{P_1/Q_1 = -6 ,\, P_2/Q_2 =-3/2}$. Here $\mathbf{4}_1$ represents the figure-eight knot while $\mathbf{5}^2_1$ represents the Whitehead link.

The state-integral is schematically written as
\begin{equation}
\mathcal{Z}_{\hbar}^\textrm{SI}(M_3;N) =\int  \frac{d^s \vec{Z}}{(2\pi \hbar)^{s/2}} \; \mathcal{I}_\hbar (\vec{Z};M_3,N) \;,
\end{equation}
and the integrand can be asymptotically expanded as
\begin{equation}
\log \mathcal{I}_\hbar (\vec{Z};M_3,N)  \qquad\xrightarrow{\; \hbar \rightarrow 0 \;} \qquad \sum_{n=0}^\infty \mathcal{W}_n \bigl( \vec{Z},\vec{\ell}_z;M_3,N \bigr)\;.
\end{equation}
Here $\vec{\ell}_z \in \mathbb{Z}^s$ comes from the branch cut structure of $\log \mathcal{I}_\hbar$ in the asymptotic limit. In particular $\mathcal{W}_0$ and $\CW_{1}$ depend on the choice of sheet in the following way:
\begin{align}
\begin{split}
&\mathcal{W}_0 (\vec{\ell}_z) = \mathcal{W}_0 (\vec{\ell}_z = \vec{0})  +2 \pi i  \vec{\ell_z} \cdot \vec{Z}\;,
\\
&\mathcal{W}_1 (\vec{\ell}_z) = \mathcal{W}_1 (\vec{\ell}_z = \vec{0})  -2 \pi i  \sum_{i=1}^2 (\vec{\ell_z})_i \;,
\end{split}
\end{align}
and $\mathcal{W}_{n\geq 2}$ are independent of the choice.  Integration of the state-integral model along a proper cycle gives the squashed three-sphere partition function of $\mathcal{T}_N[M_3]$ theory. 
The classical part $\mathcal{W}_0$ is equal to the twisited superpotential, and the Bethe-vacua of $\mathcal{T}_N[M_3]$ are obtained by extremizing it:
\begin{equation}
\mathcal{S}_{\rm BE} \bigl(\mathcal{T}_N[M_3] \bigr) =  \left\{ \vec{z} \; \middle| \; \exp \left(\partial_{Z_i} \mathcal{W}_0 (\vec{Z},\vec{\ell_z};M_3,N)\right)\bigg{|}_{Z_i = \log z_i}=1\;, \textrm{ for }i=1,\ldots, s \right\} \;.
\end{equation}
For each $\vec{z}^\alpha \in \mathcal{S}_{\rm BE} \bigl( \mathcal{T}_N[M_3] \bigr)$, there is a corresponding saddle point $\vec{Z}^\alpha$  
\begin{align}
\partial_{Z_i} \mathcal{W}_0 (\vec{Z}, \vec{\ell_z};M_3,N) =0\;, \quad i=1,\ldots s
\end{align}
for a proper choice of $\vec{\ell}_z \in \mathbb{Z}^s$. Using the 3d-3d relation \eqref{3d-3d for Bethe-vacua}, each saddle point can be identified with an $SL(N,\mathbb{C})$ flat connection (modulo tensoring with $\mathbb{Z}_N$ flat connections) on $M_3$:
\begin{align}
\vec{Z}^\alpha  \; \; \Leftrightarrow \;\; [\mathcal{A}^\alpha] \in \frac{\chi_{\rm irred}[N;M_3]}{\textrm{Hom}[\pi_1 (M_3)\rightarrow SL(N,\mathbb{C})]}\;.
\end{align}
Then, the perturbative invariants $\bigl\{ S^{\alpha}_n (M_3;N) \bigr\}$ around a flat connection $\mathcal{A}^\alpha$ can be obtained from a formal perturbative expansion of the  state-integral around the corresponding saddle point $\vec{Z}^\alpha$. The precise relation is
\begin{equation}
\label{Perturbative invariants from SI}
\mathcal{Z}_\hbar^{\rm SI} (M_3 ;N)
\quad \xrightarrow{\; \parbox{2.3cm}{\scriptsize expansion around \\ saddle point $\vec{Z}^\alpha$} \; } \quad \sqrt{\frac{\bigl| \textrm{Hom}[\pi_1(M_3)\rightarrow \mathbb{Z}_N] \bigr| }{N}} \; \exp \left( 
\sum_{n} \hbar^{n-1}S_n^{\alpha}(M_3;N)\right)\;.
\end{equation}

\paragraph{Example: \matht{M_3 = (S^3\backslash \mathbf{4}_1)_{P/Q=5}}.}
Here $\mathbf{4}_1$ denotes the figure-eight knot. 
The corresponding 3d gauge theory is given in \eqref{T[M3]-for-M3=(41)-5} and the state-integral model is given by
\begin{align}
\begin{split}
\mathcal{Z}^{\rm SI}_{\hbar} \bigl( M_3 = (S^3\backslash \mathbf{4}_1)_{5};N=2 \bigr) &= \int \frac{dZ}{\sqrt{2\pi \hbar}} \; \psi_{\hbar}(Z) \; \exp \left(  -\frac{3}{2 \hbar}   Z^2 + \frac{1}\hbar (i \pi +\hbar/2 ) Z \right) 
\\
&=  \int \frac{dZ}{\sqrt{2\pi \hbar}} \; \mathcal{I}_\hbar \bigl( Z; M_3  =  (S^3\backslash \mathbf{4}_1)_{5} \bigr)\;.
\end{split}
\end{align}
Here $\psi_{\hbar} (Z)$ is a special functional called quantum dilogarithm (QDL):
\begin{align}
\psi_\hbar (Z) := \begin{cases} \displaystyle\prod_{r=1}^\infty \frac{1-q^r e^{-Z}}{1-\tilde{q}^{-r+1}e^{-\tilde{Z}}}&\mbox{if } |q|<1\;, \\ 
\displaystyle \prod_{r=1}^\infty \frac{1-\tilde{q}^r e^{-\tilde{Z}}}{1-q^{-r+1}e^{-Z}} &\mbox{if } |q|>1 \;,\\ 
\end{cases} \label{def : Q.D.L}
\end{align}
with
\begin{align}
q:=e^{2\pi i b^2}\;, \quad \tilde{q}:=e^{2\pi i b^{-2}}\;, \quad \tilde{Z}:= \frac{1}{b^2}Z\;.
\end{align}
The asymptotic  expansion when $\hbar =2\pi i b^2 \rightarrow 0$ is given by
\begin{align}
\log\psi_{\hbar}(Z) \; \xrightarrow{\;\;\hbar \rightarrow 0\;\;} \;  \sum_{n=0}^{\infty} \frac{B_n \, \hbar^{n-1}}{n!} \; \widetilde{\textrm{Li}}_{2-n}(e^{-Z},\ell_z)\;. \label{asymptotic of QDL}
\end{align}
Here $B_n$ are the Bernoulli numbers with $B_1=1/2$. Moreover
\begin{align}
\begin{split}
&\widetilde{\textrm{Li}}_{2} (e^{-Z},\ell_z)= \textrm{Li}_2  (e^{-Z}) + 2\pi i \ell_z Z
\\
&\widetilde{\textrm{Li}}_{1} (e^{-Z},\ell_z)= \textrm{Li}_1  (e^{-Z})-2\pi i \ell_z
\\
&\widetilde{\textrm{Li}}_{2-n} (e^{-Z},\ell_z)= \textrm{Li}_{2-n}  (e^{-Z}) \quad (n\geq 2)
\end{split}
\end{align}
and $\ell_z \in  \mathbb{Z}$ is a locally-constant function of $Z$ which comes from the branch cut structure of $\textrm{Li}_{1,2}$. Using the asymptotic expansion of the QDL, we can asymptotically expand the integrand as follows:
\begin{align}
\begin{split}
&\mathcal{I}_\hbar (Z) \; \xrightarrow{\;\hbar \rightarrow 0\;} \;   \exp \left( \sum_{n=0}^\infty  \hbar^{n-1} \, \mathcal{W}_n (Z,\ell_z) \right)\;, \quad  \textrm{where}
\\
&\mathcal{W}_0 (Z;\ell_z)  = \textrm{Li}_2  (e^{-Z}) + i \pi (1+2 \ell_z)Z- \frac{3}2 Z^2\;,
\\
&\mathcal{W}_1 (Z;\ell_z)  = \textrm{Li}_1  (e^{-Z})-2\pi i \ell_z + \frac{1}2 Z\;, \qquad\quad\dots
%\\
%&\ldots.
\end{split}
\end{align}
Then $\mathcal{W}_0$ is the twisted superpotential and the Bethe vacua are given by the following algebraic equation in $z:=e^Z$:
\begin{align}
\exp \left( \partial_Z \mathcal{W}_0 \right) =1\; \quad \Rightarrow  \quad \frac{1-z}{z^4}=1 \;.
\end{align}
For each Bethe vacuum $z^\alpha$, there is a corresponding $Z^{\alpha}$ with a proper choice of $\ell_z^\alpha$ such that
\begin{align}
\partial_Z \mathcal{W}_0 \Big|_{Z = Z^\alpha,\, \ell_z = \ell_z^\alpha} = 0\;.
\end{align}
The Bethe vacua are in  one-to-one correspondence with irreducible $SL(2,\mathbb{C})$ flat connections on $M_3$.  We can perturbatively expand the state-integral around each  saddle point $Z^\alpha$ as follows:
\begin{align}
&\frac{1}{\sqrt{2}} \exp \left( \sum_{n=0}^\infty \hbar^{n-1} S_n^\alpha (M_3;N=2) \right) 
= \int \frac{d (\delta Z)}{\sqrt{2\pi}} \mathcal{I}_\hbar \left(Z^\alpha +\hbar^{1/2}\delta Z;M_3 \right) 
\nn \\
&=\exp \left(\sum_{n=0}^\infty \mathcal{W}_n (Z^\alpha) \right)  \int \frac{d (\delta Z)}{\sqrt{2\pi }} \exp \left(- \frac{1}2 \Pi^\alpha (\delta Z)^2  + \sum_{n=1}^\infty \sum_{1\leq m \leq n+2; m-n \in 2\mathbb{Z}} \hbar^{n/2} C^\alpha_{n,m} (\delta Z)^m\right)
\nn \\
&=\exp \left(\sum_{n=0}^\infty \mathcal{W}_n (Z^\alpha) \right)  \int \frac{d (\delta Z)}{\sqrt{2\pi }} \exp \left(- \frac{1}2 \Pi^\alpha (\delta Z)^2   \right) \left(1+ \sum_{n=1}^\infty \sum_{m=1}^{3n} \hbar^{n}D_{n,m}^\alpha (\delta Z)^{2m} \right)
\nn \\
&=\exp \left( - \frac{1}2 \log \Pi^\alpha + \sum_{n=0}^\infty \mathcal{W}_n (Z^\alpha) \right) \left(1+ \sum_{n=1}^\infty \sum_{m=1}^{3n} \hbar^{n}D_{n,m}^\alpha G^\alpha_m \right)\;.
\end{align}
Here we defined:
\begin{align}
\begin{split}
&\Pi^\alpha := -\partial^2_Z \mathcal{W}_0 \bigg{|}_{Z = Z^\alpha} \quad \textrm{(propagator)}\;,
\\
&C^\alpha_{n,m} := \frac{1}{m!}\partial^m_Z \mathcal{W}_{\frac{n-m}2+1} \bigg{|}_{Z = Z^\alpha}\quad \textrm{(vertices)}\;,
\\
&D^\alpha_{n,m}:=  \textrm{coefficient of $\hbar^n (\delta Z)^{2m}$ in } \exp \left(\sum_{n=1}^\infty \sum_{1\leq m \leq n+2; m-n \in 2\mathbb{Z}} \hbar^{n/2} C^\alpha_{n,m} (\delta Z)^m\right)\;,
\\
& G_m^\alpha :=  \sqrt{\Pi^\alpha} \int \frac{d (\delta Z)}{\sqrt{2\pi}} \exp \left(- \frac{1}2 \Pi^\alpha (\delta Z)^2 \right) (\delta Z)^m = \partial^{2m}_\mu \exp \left(\frac{1}2 (\Pi^\alpha)^{-1} \mu^2 \right)\bigg{|}_{\mu=0}\;. \label{perturbative computation}
\end{split}
\end{align}
The perturbative coefficients $\bigl\{S_n^{\alpha}(M_3 = (S^3\backslash \mathbf{4}_1)_{P/Q=5}; N=2 ) \bigr\}$ computed as above are given in \eqref{Sn for (41)5}  up to $n=5$. 

\paragraph{Example: \matht{M_3 =\textrm{Weeks manifold} = (S^3 \backslash \mathbf{5}^2_1 )_{P_1/Q_1 =- 5,\, P_2/Q_2 = -5/2}}.}
Here $\mathbf{5}^2_1$ denotes the Whitehead link. The corresponding 3d gauge theory is given in \eqref{3d-theory-for-weeks}. The corresponding state-integral model is given by
\begin{align}
\mathcal{Z}^{\rm SI}_{\hbar} \left( \textbf{Weeks}\right) = \int \frac{dZ}{\sqrt{2\pi \hbar}} \; \psi_{\hbar}(Z) \; \exp \left(  -\frac{Z^2}{ \hbar}   \right) \;.
\end{align}

\paragraph{Example: \matht{ M_3 =(S^3 \backslash \mathbf{5}^2_1 )_{P_1/Q_1 =- 6,\, P_2/Q_2 = -3/2}}.}
Here $\mathbf{5}^2_1$ denotes the Whitehead link. The corresponding 3d gauge theory is given in \eqref{3d-theory-for-3rd}, and the state-integral model is
\begin{align}
\mathcal{Z}^{\rm SI}_{\hbar} \bigl( M_3 =(S^3 \backslash \mathbf{5}^2_1 )_{P_1/Q_1 =- 6,\, P_2/Q_2 = -3/2} \bigr)= \int \frac{dZ}{\sqrt{2\pi \hbar}} \; \psi_{\hbar}(Z) \; \exp \left(  -\frac{Z^2}{ 2\hbar}   \right) \;. 
\end{align}

\bibliographystyle{ytphys}
\bibliography{ref_M5entropy}
\end{document}